\documentclass[apj]{emulateapj}
\usepackage{apjfonts}

\newcommand{\bx}{\mbox{\boldmath $x$}}
\newcommand{\bl}{\mbox{\boldmath $l$}}
\newcommand{\bk}{\mbox{\boldmath $k$}}
\newcommand{\bov}{\mbox{\boldmath $v$}}
\newcommand{\bog}{\mbox{\boldmath $g$}}
\newcommand{\bnabla}{\mbox{\boldmath $\nabla$}}

\def\s{{\rm\,s}}
\def\erg{{\rm\,erg}}

\def\cm{{\rm\,cm}}

\def\kpc{{\rm kpc}}

\def\K{{\rm\,K}}
\def\yr{{\rm\,yr}}

\shorttitle{Supernova-Driven ISM}
\shortauthors{Joung \& Mac Low}

\begin{document}

\title{Turbulent Structure of a Stratified Supernova-Driven Interstellar Medium}

\author{M.~K.~Ryan~Joung\altaffilmark{1,2} and Mordecai-Mark~Mac~Low\altaffilmark{1,2}}
\altaffiltext{1}{Department of Astronomy, Columbia University, 550 West 120th Street, New York, NY~10027; moo@astro.columbia.edu}
\altaffiltext{2}{Department of Astrophysics, American Museum of Natural History, 79th Street at Central Park West, New York, NY~10024; mordecai@amnh.org}

\begin{abstract}
To study how supernova feedback structures the turbulent interstellar medium, we 
construct 3D models of vertically stratified gas stirred by discrete 
supernova explosions, including vertical gravitational field and parametrized 
heating and cooling. The models reproduce many observed characteristics of the 
Galaxy such as global circulation of gas (i.e., galactic fountain) and the 
existence of cold dense clouds in the galactic disk. Global quantities of the model 
such as warm and hot gas filling factors in the midplane, mass fraction of 
thermally unstable gas, and the averaged vertical density profile are compared 
directly with existing observations, and shown to be broadly consistent. 
We find that energy injection occurs over a broad range of scales. There is no 
single effective driving scale, unlike the usual assumption for idealized models of 
incompressible 
turbulence. However, $>$90\% of the total kinetic energy is contained in wavelengths 
shortward of 200 pc. The shape of the kinetic energy spectrum differs substantially 
from that of the velocity power spectrum, which implies that the velocity structure 
varies with the gas density. Velocity structure functions demonstrate that the 
phenomenological theory proposed by Boldyrev is applicable to the medium.
We show that it can be misleading to predict physical 
properties such as the stellar initial mass function based on numerical simulations 
that do not include self-gravity of the gas. Even if all the gas in turbulently Jeans 
unstable regions in our simulation is assumed to collapse and form stars in local 
freefall times, the resulting total collapse rate is significantly lower than the 
value consistent with the input supernova rate. Supernova-driven turbulence 
inhibits star formation globally rather than triggering it.
\end{abstract}


\section{Introduction}

The interstellar medium (ISM) is observed to be turbulent. The electron density 
spectrum in the warm ionized medium shows a nearly Kolmogorov spectrum over at 
least six orders of magnitude in length (Armstrong et al. 1995). Supposing the small 
density perturbations measured to behave as a passive scalar, this can be taken as 
evidence for a Kolmogorov velocity spectrum in that gas. 
Giant molecular clouds, which contain most of the molecular gas, show 
supra-thermal emission linewidths (Zuckerman \& Palmer 1974), indicating 
the presence of supersonic random motions. 

What drives this turbulence?
Massive stars are a major structuring agent in the ISM (McCray \& Snow 1979). Once 
formed inside molecular clouds, they influence the surrounding medium via intense ionizing 
radiation and stellar winds until they end their lives in catastrophic explosions 
as supernovae, heating and stirring the ISM throughout their lives. Among these 
energetic processes, supernovae (and their collective effect, 
superbubbles) are likely to be the dominant contributor to the observed supersonic 
turbulence (Norman \& Ferrara 1996; see Mac Low \& Klessen 2004 for a recent 
review). This turbulence changes the balance between pressure and gravity on a 
wide range of scales, from galactic (Boulares \& Cox 1990) to sub-cloud scales 
(e.g., Larson 1981), and in turn may regulate subsequent star formation, hence 
the term ``feedback.'' With strong enough feedback, a galactic-scale superwind 
may be driven out of the galaxy (Heckman et al. 2000, 2001).

Our understanding of how the radiative and hydrodynamic feedback affects spatio-temporal 
structure of the ISM and star formation is only qualitative. It remains 
difficult to incorporate the feedback into a consistent theory of star formation 
and the ISM, partly because star formation is an inherently complex phenomenon 
that involves an interplay between gravity, magnetohydrodynamics, and thermodynamics 
of compressible gas. Yet it is a crucial physical process for 
correct predictions of, e.g., the thickness, disk stability, and star formation 
rate of galaxies. Many of the serious problems that cosmological models are faced 
with today are also believed to be associated with such feedback.\footnote{The standard 
CDM paradigm predicts basic properties of the 
luminous component of matter on galactic and sub-galactic scales which are seriously 
inconsistent with observations. Outstanding astrophysical problems include incorrect 
predictions for the shape of the galaxy luminosity function and for the fraction 
of gas that should have cooled to form galaxies (the overcooling problem) (Somerville 
\& Primack 1999; Cole et al. 2000).}

Since no analytic theory based on first principles exists for turbulence in a 
compressible medium (see, however, a heuristic model by Sasao 1973), most previous 
work has relied heavily on numerical methods. 
Recent numerical simulations in periodic boxes yielded key insights 
into the role of turbulence in star formation. First, supersonic turbulence 
in an isothermal or an adiabatic medium was found to decay quickly, i.e. within a 
sound crossing time, whether the medium was magnetized or not (Stone et al. 1998; 
Mac Low et al. 1998; Padoan \& Nordlund 1999), unless it was constantly 
stirred by an external 
source of driving (Mac Low 1999). Second, supersonic turbulence delays or sometimes 
prevents collapse (Klessen et al. 2000, hereafter KHM00), and is hence 
thought to be responsible for the observed low overall star formation efficiencies 
(V\'{a}zquez-Semadeni et al. 2005). Overall, turbulence impedes 
collapse in unstable regions. Third, in contrast to the second point, it also creates 
converging flows 
that may enhance density and thus promote collapse (Ballesteros-Paredes et al. 
1999). Even when turbulence can support a cloud globally against gravitational 
collapse, it produces density enhancements that allow local collapse in both 
non-magnetized (KHM00) and magnetized media (Heitsch, Mac Low, \& Klessen 
2001, hereafter HMK01). However, the net effect of turbulence is to inhibit collapse 
(Mac Low \& Klessen 2004). 

In some of these previous simulations, turbulence was seeded as the initial 
condition and left to decay for the rest of the run (Porter et al. 1992; 
Porter et al. 1998; Stone et al. 1998; Mac Low et al. 
1998; Padoan \& Nordlund 1999). In other simulations, turbulence was driven constantly 
but the driving was included only in a general way
as ubiquitous Fourier forcing, i.e., a random forcing in a prescribed 
narrow range of wavenumbers applied uniformly to the box (Mac Low 1999; 
KHM00; HMK01). These models only partly represent the real ISM, where 
forced and decaying regimes may coexist (Avila-Reese \& V\'{a}zquez-Semadeni 2001) and 
where an appropriate forcing function is probably 
broad-band (Norman \& Ferrara 1996). It has been demonstrated that the slope of the power spectrum for 
the forcing function profoundly affects the statistics of turbulent fluctuations 
(Bonazzola et al. 1987; V\'{a}zquez-Semadeni \& Gazol 1995; Biferale et al. 2004). 
Hence, it is of great interest to determine the wavelength distribution of the 
kinetic energy in a more realistic medium driven by \textit{discrete physical space 
forcing}.

Our goal is to examine physical characteristics of the ISM driven 
by realistic, discrete physical space forcing in 3D. In particular, we study 
density and velocity structures and determine the kinetic energy power spectrum. 
If, as some previous works suggest, the interaction of blast waves from supernovae 
structures the interstellar gas, we must resolve the individual explosions in 
order to correctly reproduce the main statistical properties of the ISM.
For an accurate simulation of supernova driving, a wide range of length scales 
from small clouds ($\sim$1 pc) to at least the ``driving scale'' (using the language 
derived from incompressible turbulence research) must be resolved. In addition, 
to predict the cloud structure, the model must take into account appropriate 
heating and radiative cooling processes.

There have been modeling efforts toward more realistic supernova driving by 
including its explosive nature. Rosen, Bregman, \& Norman (1993), Rosen \& 
Bregman (1995), and Wada \& Norman (2001) performed two-dimensional hydrodynamic 
simulations for the ISM in a galactic disk including feedback from 
supernovae and/or stellar winds. Results from two-dimensional models 
must be interpreted with caution, since turbulence behaves differently in 2D 
than in 3D: in 2D, an inverse cascade of energy occurs towards large scales, 
while the enstrophy, the square of vorticity, cascades to small scales (Kraichnan 
1967; Frisch 1995). More recently, three-dimensional models have been 
studied by several groups. These authors find that blast waves from 
supernovae sweep up the ISM into relatively thin shells that collide with 
one another to form cold dense clouds. 
However, none of the goals raised above has been adequately addressed. 
Korpi et al. (1999) used resolution of 
10 pc, insufficient to follow the details of turbulent interactions, and only 
covered (0.5 kpc)$^2 \times$ 2 kpc, although they did include magnetic field 
and galactic shear. Their main interest was to simulate a turbulent galactic 
dynamo. Slyz et al. (2005) included self-gravity of the gas but suffer from low 
spatial resolution ($\ge$10 pc) and the absence of vertical density stratification. 
Kim et al. (2001) and Mac Low et al. (2005) also included magnetic field, but not 
galactic stratification. Avillez (2000) and Avillez \& Berry (2001) used a stratified 
model similar to ours, but their studies focused on the properties of a Galactic 
fountain in the Milky Way. All the aforementioned 3D models employed minimum 
gas temperatures $T_{min} = 100$--310 K, so no cold gas could form. The difference 
between Avillez \& Breitschwerdt (2004a, b) and the present work is more subtle, and 
will be discussed in \S\ \ref{midplane}.

We simulate a small patch of our Galaxy, (0.5 kpc)$^2$ in area. The computation 
box is elongated in $z$ and extends from $-5$ kpc to $+5$ kpc to study the vertical 
structure. We choose the length 
scales in our simulations ($2$ pc $\le l \le$ $10$ kpc) to lie between those 
of star-forming molecular clouds and those of large-scale galactic outflows 
(Heckman et al. 2000, 2001). Our box size represents an optimal choice to 
improve our understanding of this enigmatic yet relatively unexplored regime. 
Expanding shells and their interactions are well-resolved at this scale. 
We attempt to build local, 
high-resolution models of the ISM based on which subgrid models for turbulent pressure 
can be developed, that will provide insight into how supernova feedback should 
be treated in global, cosmological simulations. We will continue to pursue this problem 
in a companion paper (M. Joung \& M. Mac Low 2005, in preparation; hereafter Paper II). 
Here, in Paper I, we 
describe our numerical model (\S\ \ref{model}) and present the basic results (\S\ 
\ref{results}). In \S\ \ref{clouds}, we explore where gravitational 
collapse would occur in the presence of explosion-driven turbulence. Finally, 
we summarize our findings and discuss their implications in \S\ \ref{discuss}.

\section{The Model}
\label{model}
\subsection{Basic Features}
\label{features}

Our simulations are performed using Flash (Fryxell et al. 2000), an 
Eulerian astrophysical hydrodynamics code with adaptive mesh refinement 
(AMR) capability, developed by the Flash Center at the University of 
Chicago. It solves the Euler equations using the piecewise-parabolic 
method (Colella \& Woodward 1984) to handle compressible flows with shocks. 
For parallelization, the Message-Passing Interface library is used; 
the AMR is handled by the PARAMESH library.

We set up a stratified gas at 1.1$\times$10$^4$ K initially in hydrostatic 
equilibrium. The surface mass density of the gas, $\Sigma_{gas} = 
7.5$ M$_{\odot}$ pc$^{-2}$.  
The computation box contains a volume of (0.5 kpc)$^2 \times$(10 kpc), 
elongated in the vertical direction. At 1.95 pc resolution, 
it effectively contains 256$^2 \times$5120 zones. We modify the 
refinement and derefinement criteria so that it becomes gradually 
more difficult to refine (and easier to derefine) with increasing 
distance from the midplane ($z=0$). In practice, the inner $\sim$600 pc near 
the midplane ends up refined to the maximum level. 
This allows us to capture the vertical flow, while not unnecessarily 
resolving low-density regions far from the disk. For our fiducial 
model, the number of dynamically allocated zones is only 7--8 \% of 
that in a single mesh code with the same maximum resolution.

The gravitational potential of Kuijken \& Gilmore (1989) is 
employed. It yields the vertical gravitational acceleration 
\begin{equation}
\label{gofz}
g(z) = -\frac{a_1 z}{\sqrt{z^2+z_0^2}} - a_2 z \; ,
\end{equation}
where $a_1 = 1.42 \times 10^{-3}$ kpc Myr$^{-2}$, $a_2 = 5.49 
\times 10^{-4}$ Myr$^{-2}$, and $z_0 = 0.18$ kpc. The two terms 
on the RHS of the equation represent contributions from a stellar disk 
and a spherical dark halo. 
It is static in time; self-gravity of the inhomogeneous, evolving gas 
is not included. Outflow boundary conditions are used on the upper and 
lower surfaces parallel to the Galactic plane, while periodic 
boundary conditions are used elsewhere.

For each supernova explosion, we add thermal energy $E_{sn} 
= 10^{51}$ ergs in a small sphere whose radius varies as a 
function of the local density. The radii are chosen such that 
radiative losses inside the spheres are negligible in the first 
few timesteps after the explosions. The details are laid out in \S\ 
\ref{sne} below. We assume a constant overall rate of supernova 
explosions, and treat it as an input parameter. This assumption of 
constant supernova rate is justified by the fact that dynamically 
important variations in the supernova rate occur on $\sim$1 Gyr time 
scale, far longer than the duration of our simulations.
 
In addition, each supernova produces $\sim$500 metal particles that 
evolve passively following the gas component. We tag each particle with 
a unique number and follow its position, velocity, and thermal properties 
of the gas (density, temperature, and pressure) in the cell that contains 
the particle. These variables are recorded at every timestep for all 
particles. This procedure 
enables us to trace the thermal history of metal particles and to 
estimate their escape fraction from the galaxy. 

We assume that a fixed fraction ($15/31$ in the fiducial model; \S\ 
\ref{sne}) of the supernovae are closely correlated in 
space as a way of simulating superbubbles. The remaining 
explosions have random positions scattered through the galaxy mass 
to represent field supernovae. 

\begin{figure}
\epsscale{1.0}
\includegraphics[scale=0.42]{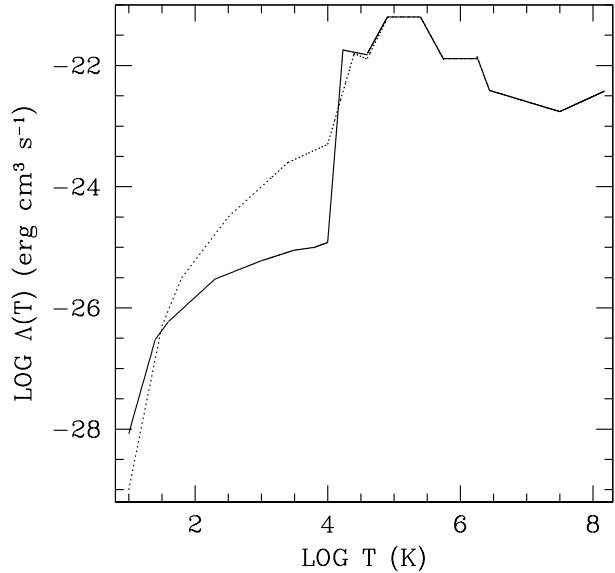}
\caption{Radiative cooling function for an optically thin plasma with 
$Z/Z_{\odot}=1$ approximated as a piecewise power law. The cooling curve 
is parametrized following Dalgarno \& McCray (1972) with an ionization 
fraction of $10^{-2}$ at $T \le 2 \times 10^4$ K and Sutherland \& Dopita (1993) 
at $T \ge 2 \times 10^4$ K. Equilibrium ionization is assumed. For comparison, the 
cooling curve from Spaans \& Norman (1997) is shown in dotted line. 
\label{fig1}}
\end{figure}

We solve the standard hydrodynamic equations, namely the continuity 
equation, the momentum equation, and the energy equation in a background 
gravitational field. Here we write 
only the energy equation in our model, which is modified to include 
supernova heating. In conservative form, 
\begin{equation}
\label{dedt}
\frac{\partial \, \rho E}{\partial t} + \bnabla \cdot [\, (\rho E + P) \, \bov \,] 
= \rho \, \bov \cdot \bog - n^2 \, {\cal L} + S \; ,
\end{equation}
where $\rho$, $P$, $\bov$, and $T$ denote density, pressure, velocity, 
and temperature of the gas; the specific total energy $E$ is the sum 
of internal energy $\varepsilon$ and kinetic energy per unit mass: 
$E \equiv \varepsilon + |\bov|^2/2$. Gas pressure is related to density 
via the relation $P = (\gamma - 1) \rho \varepsilon$ with the adiabatic 
index $\gamma = 5/3$. The gravitational acceleration $\bog = g(z)\hat{z}$ 
given by equation (\ref{gofz}). The net cooling rate per unit volume 
$n^2 \, {\cal L} = n^2 \, \Lambda(T) - n \, \Gamma(z)$, where $n$ is 
the number density of gas. 
For the cooling function, $\Lambda(T)$, radiative cooling appropriate 
for an optically thin plasma with $Z/Z_{\odot} = 1$ (cosmic abundance) 
is included assuming equilibrium ionization (Sutherland \& Dopita 1993). 
It is displayed in Figure \ref{fig1}, approximated as a piecewise power law. 
When the cooling function is expressed as $\Lambda_i(T) \propto T^{\beta_i}$, 
the gas is thermally stable where $\beta_i \ge 1$. For the particular cooling curve 
that we adopted, the gas is thermally stable for $10\K \leq T < 40\K$ and 
$10^4\K \leq T < 1.7\times10^4\K$, and marginally stable for $40\K \leq T 
< 200\K$. 
The portion of the curve at low temperatures ($T \le 2\times10^4$ K) 
is adopted from Dalgarno \& McCray (1972) assuming an ionization fraction 
of $10^{-2}$. The cooling rate is most uncertain in this regime since 
the approximation of thermal equilibrium  is usually invalid. Also shown 
in dotted line in Figure \ref{fig1} is the cooling curve from Spaans \& 
Norman (1997), as reproduced in Wada \& Norman (2001; see their Figure 1). 
The exact forms for the two heating terms in our 
model, the diffuse heating function $\Gamma(z)$ and the impulsive heating 
from local supernova explosions $S(\bx, t)$, are given in \S\S\ \ref{dheat} 
and \ref{sne}, respectively. 

In dense regions, the cooling time $t_{cool} \equiv (\gamma-1)^{-1}kT/n\Lambda$ 
can be shorter than the 
Courant timestep $\Delta t$. FLASH, as a default, applies heating and 
cooling terms in sequence using an explicit method for updating temperatures. 
We have combined the separate heating and cooling routines so that temperature 
is updated only once in each timestep, and used an implicit method 
whenever the \textit{net} cooling time $t_{net} \equiv (\gamma-1)^{-1}kT/n 
{\cal L} \leq 0.1 \Delta t$. Thus, temperatures for high density ($n \gtrsim 
10 \cm^{-3}$) gas are computed using an iterative procedure, in our case, 
Brent's method (Press et al. 1992).

\subsection{Diffuse Heating}
\label{dheat}

It is crucial to choose the right diffuse heating rate. Without this heating, 
high density gas will simply cool down to $T_{min}$ within a cooling time 
given by $t_{cool}$. In other words, no thermal equilibrium will exist above 
$T_{min}$. Previous numerical models adopted a diffuse heating rate $\Gamma_{hs}$ 
which balances the vertical gravity $g(z)$ on the gas to yield hydrostatic equilibrium at a 
uniform prescribed temperature $T_{init}$ (Mac Low et al. 1989; Avillez \& 
Breitschwerdt 2004a). This heating rate is computed so that the net cooling rate in 
equation (\ref{dedt}) is zero for all $z$: $n \, \Gamma(z) = n^2 \, 
\Lambda(T_{init})$, for the hydrostatic equilibrium density profile given in 
equation \ref{isot} below. If the gas were isothermal, the cooling term would dominate 
where $\rho>\rho_{hs}$, while the diffuse heating term would dominate where 
$\rho<\rho_{hs}$. 

In contrast, 
photoelectric heating, i.e. photoemission of UV-irradiated dust grains (Bakes 
\& Tielens 1994), has long been thought to be the dominant form of heating 
for the neutral component of the ISM, i.e., the cold neutral medium (CNM) and the 
warm neutral medium (WNM)
(Wolfire et al. 1995). It has the form $\Gamma_{pe} = 1.0 \times 10^{-24} 
\epsilon G_0 \erg \s^{-1}$, where the heating efficiency $\epsilon \approx 
0.05$ and $G_0$ is the incident far-ultraviolet field normalized 
to Habing's (1968) estimate of the local interstellar value (Gerritsen \& Icke 
1997). We adopt Draine (1978)'s value $G_0 = 1.7$. 

Ionizing radiation from OB stars is an additional and a significant heating 
source. Abbott (1982) estimates a total integrated luminosity $\dot{e} = 1.5 
\times 10^{-24}$ erg s$^{-1}$ cm$^{-3}$ in the disk of the Milky Way. 
Since only 20\% of all available UV photons go into heating interstellar gas 
(Abbott 1982), the photoionization heating rate 
$\Gamma_{pi} \approx 3.5 \Gamma_{pe}$. Much of this heating goes to the warm 
ionized medium. We do not, however, include it in our diffuse heating rate.

Initially, the gas is assumed to be isothermal. We solve for the hydrostatic equilibrium 
profile of density for the given gravitational potential (see equation \ref{gofz}):
\begin{equation}
\label{isot}
\rho_{hs}(z) = \rho_0 \, \exp \left(-\frac{a_1 \rho_0}{P_0}\left(\sqrt{z^2+z_0^2}-z_0\right)
-\frac{a_2 \rho_0}{2 P_0}z^2\right) \; ,
\end{equation}
where $\rho_0$ and $P_0$ are the gas density and pressure at the midplane; 
$P_0 = (\rho_0/\mu m_H) \, k \, T_0$ where $T_0$ denotes 
the isothermal temperature of the gas. Note that the profile is narrower 
than a gaussian by the exponential factor due to a stellar disk (the term 
containing $a_1$). This profile, $\rho_{hs}(z)$, would describe the vertical 
density distribution if the only form of support were thermal pressure. 
We evolve the systems for sufficiently long periods of time so that the 
memory of the initial condition is completely lost.

Attempting to balance radiative cooling and diffuse heating of gas 
at our chosen initial gas temperature $T_{init} = 1.1$$\times$10$^4 \K$ leads 
to $\Gamma_{hs} \approx 30 \Gamma_{pe}$. We believe that $\Gamma_{hs}$, 
as used in previous models, is unrealistically high. It is the combination 
of proper (in fact, much lower) diffuse heating and supernova heating that 
together should balance radiative cooling.

For the diffuse heating rate in our fiducial model, we adopt $\Gamma_{pe}$, 
a value consistent with Wolfire et al. (1995). Note that our value of 
$\Gamma$ is lower than that in Avillez \& Breitschwerdt (2004a) by a factor of 
$\sim$18 (their $T_{init}=9.0 \times 10^3 \K$. The heating rate coefficient 
$\Gamma$ is assumed to be independent of 
gas density; according to Wolfire et al. (1995), photoelectric heating weakly 
depends on density: $\Gamma_{pe} \propto \rho^{0.2}$. 

The diffuse heating is applied to gas with $T \leq 
T_{ion}$, where $T_{ion} = 2\times10^4$ K is the temperature at 
which hydrogen is fully thermally ionized. It is also assumed to 
decline exponentially in $z$ as $\Gamma \propto \Gamma_0 \, e^{-z/H_{dh}}$. 
This exponential factor is a measure for the escaping ionizing radiation 
from the disk (Domg\"orgen \& Mathis 1994; Dove \& Shull 1994; Bland-Hawthorn 
\& Maloney 1999, 2002), which should 
in principle be treated in a self-consistent way via a radiative transfer 
code (Fujita et al. 2003; Wood et al. 2004). Existing observations 
are unable to provide a definite value of $H_{dh}$ for Milky Way type galaxies. 
In our model, $H_{dh}$ is set to 300 pc so that $\sim$4 \% 
of the ionizing radiation leaks out of the disk and into the halo, heating 
diffuse ionized gas there (Domg\"orgen \& Mathis 1994). At high $|z|$, 
a minimum value $\Gamma_{min}=10^{-5}\Gamma_0$ was used. Note that the prescribed 
functional form for $\Gamma(z)$ precludes the possibility of obtaining thermal 
equilibrium for the initial gas distribution at all heights.

Our heating function $\Gamma$ 
varies as a function of $z$ only and is thus uniform at a given height 
in the disk (e.g., Gerritsen \& Icke 1997). In reality, the diffuse 
heating rate is extremely non-uniform in space and time; it fluctuates by 
1--2 orders of magnitude, and this large fluctuation in the heating rate 
may be responsible for the conversion of CNM to WNM and vice versa (Parravano 
et al. 2003; Wolfire et al. 2003). 
Thus, our use of constant diffuse heating rate is a simplification, 
justified in part if non-uniformities created by far more energetic 
supernova explosions dominate the thermal energy budget of the gas. 

\subsection{Adding Supernova Explosions}
\label{sne}

One major obstacle that previous large-scale models including supernova feedback 
have faced was that when thermal energy $E_{sn} = 10^{51}$ erg was added 
locally into a sphere, most of the energy was radiated away too 
quickly. This cooling happens within one or two timesteps after the explosion, 
i.e., before blast waves form and start sweeping up the surrounding medium, 
converting the initially thermal energy into kinetic energy. As a result, adding 
impulsive heating in their models did not heat the medium in any significant way 
either thermally or dynamically (Katz 1992). 
To avoid this problem, each explosion should occur in such a way that no 
significant amount of energy is lost radiatively before the expanding sphere 
forms a strong 
blast wave and the interior density drops and relaxes to the Sedov profile. 
In order to achieve this effect, Thacker \& Couchman (2001), for example, used an 
artificial time delay in cooling within the initial explosion sphere. 
Different algorithms such as assigning the explosion energy to fluid parcels 
as pure kinetic energy also have problems (Navarro \& White 1993). The fundamental cause of 
this problem is the lack of resolution in these models: relevant 
length scales for the energy source, in this case supernova explosions, lie 
below their resolution limits especially in the early phase of the expansion.
In many previous models, an ad hoc thermalization efficiency of supernova 
energy, ranging from $10^{-1}$ (Navarro \& White 1993) to $10^{-4}$ (Hernquist 
\& Mihos 1995), was assumed. Because we track the thermal and 
dynamical evolution of blast waves from supernovae explicitly, our model 
does not suffer from such limitations.

Explosion radii in our model are determined so that the enclosed gas 
mass within each sphere $M_{exp} = 60 \, M_{\odot}$ initially. In our fiducial 
model, this is less than $2 \times 10^{-5}$ of the total gas mass in the 
computation box. We neglect the mass ejected by each supernova, i.e., 
no direct exchange of mass is assumed between stars and the interstellar gas. The 
radii usually vary between $\sim$7 and $\sim$50 pc. These initial spheres take 
up only a small fraction of the total volume. The mean local density inside 
the sphere is sufficiently low and the timestep is sufficiently short 
that no delay in cooling is necessary. Inside an explosion sphere, the 
density is redistributed uniformly to $\rho_{exp} = 3 M_{exp}/4\pi r_{exp}^3$, 
and then a thermal energy $E_{sn}$ 
is injected evenly into the sphere. The timestep $\Delta t \approx 10^3$ yr. 
The value of $M_{exp}$ represents a compromise in the sense that too small 
mass leads to too high cooling rates at $T > 10^8$ K (due to bremsstrahlung) 
and too few zones within the explosion radius. On the other hand, when the 
enclosed mass is too large, the initial cooling rates may be again too high if 
the injected thermal energy heats the sphere only to temperatures $T \lesssim 10^6$ 
K. Also, the initial explosion sphere may occupy too much volume 
and thereby introduce artificial changes in the dynamics. 

As a preliminary test, we compared the time evolution of 
thermal and kinetic energies of a single supernova exploding in a uniform 
background medium (at various densities) with high resolution 
1D simulations by Cioffi et al. (1988). At 2 pc resolution, 
their equation 3.15 provides a good fit for the energy evolution.

The observed supernova frequency for the Galaxy is 1/330 yr$^{-1}$ for Type I and 
1/44 yr$^{-1}$ for Type II supernovae (Tammann et al. 1994). 
We normalize this to the area of our computation box so that the rates per unit 
area are 4.0 Myr$^{-1}$ kpc$^{-2}$ and 30.0 Myr$^{-1}$ kpc$^{-2}$. 
In the vertical direction, the frequency of supernovae is assumed to decline 
exponentially. 
The scale heights for Type I and Type II SNe are 325 pc and 90 pc, 
respectively (Heiles 1987; Miller \& Scalo 1979).

It is known that spatial and temporal clustering of supernovae significantly 
affects their impact on the ISM (McCray \& Snow 1979). However, most numerical simulations so 
far either completely ignored this correlation or treated clustered supernovae 
simply as one large constant luminosity wind bubble (Mac Low et al. 1989; 
Strickland \& Stevens 2000). Notable exceptions are 
Korpi et al. (1999), who used statistical clustering of supernovae by introducing 
a density threshold for explosion sites, and Avillez (2000) and subsequent work, 
who assumed that 60\% of all supernovae occurred at previous explosion 
sites. No size distribution of superbubbles was specified by these models. 
Observations indicate that superbubbles (or stellar clusters) follow a 
power-law distribution of the form $dN_{\rm B} \propto n_*^{-2} \, dn_* \;$ 
where $n_*$ is the number of supernovae between 
a lower cutoff $n_{min}$ and an upper cutoff $n_{max}$, and $dN_{\rm B}$ is 
the number density of 
superbubbles having a total number of supernovae between $n_*$ and 
$n_*+dn_*$ (Kennicutt et al. 1988; McKee \& Williams 1997; 
Clarke \& Oey 2002). Clarke \& Oey (2002) point out that this 
distribution maximizes the mean volume occupied by each supernova. 
Interestingly, the $n_*^{-2}$ distribution places equal 
number of supernovae in any given decade in $n_*$. This is consistent with 
the large variations in the number of massive stars in clusters. 

For 3/5 of Type II supernovae, we explicitly account for a range of superbubble 
sizes in terms of the total number of supernovae that they contain, adopting the 
$n_*^{-2}$ power-law distribution. We fix $n_{min}=7$. To choose $n_{max}$, we 
implement the following simple procedure. According to Kennicutt et al. (1988), 
there are 6--9 associations with 3500--7000 supernovae in the entire galaxy. 
Assuming the $n_*^{-2}$ distribution, we choose $n_{max}$ so that there are about 
$A_{gal}/A_{box}$ associations with $n \in$ [$n_{max}/2$, $n_{max}$] 
supernovae in the Galaxy. If we adopt $A_{gal}=\pi(10$ kpc$)^2$ and $A_{box}=(0.5$ kpc$)^2$, 
$n_{max} \approx 40$. In our model, all supernovae that belong to a particular 
superbubble explode at the same location, an approximation justified by the 
fact that once a stellar wind bubble forms, 
all subsequent supernovae explode inside it (Mac Low \& McCray 1988). All 
superbubbles are assumed to have lifetimes of $t_{SB}=40$ Myr, 
approximately the age of the least massive B star 
(8$M_{\odot}$) when it explodes. We also assume that the given number of supernovae 
are equally spaced in time over $t_{SB}$. During the first 5 Myr of each 
superbubble's lifetime, we include a constant mechanical luminosity wind originating 
from 4 pc radius source region, mimicking 
stellar wind from the O stars. Shull \& Saken (1995) argue that such early heating from 
stellar winds may accelerate shell growth and velocity evolution, compared to 
constant-luminosity models. The integrated luminosity of stellar wind heating is 
taken to be $0.14 (n_* E_{sn})$ (Ferri\`{e}re 1995). Mass outflows from stellar winds are 
not included. Isolated supernovae, which account 
for the remaining explosions, have random positions scattered throughout the disk. 
Since the progenitor stars drift apart with velocities 
$\sim$5 km s$^{-1}$, they should lie out of their parent clouds 
by the time they explode, usually $\gtrsim$ 10 Myr (for Type IIs) and $>$ 1 Gyr 
(for Type Ia's) after they form. 

\subsection{Missing Physics}

Aside from the limited spatial resolution of our simulations, our 
numerical model neglects several physical 
processes that are known to participate in interstellar cloud evolution 
and star formation. For example, since no self-gravity is included in our model, 
unstable regions do not collapse to form stars; instead, they just remain 
dense as a whole, so it is harder for shocks to destroy the clouds.
A more complete model in the future should also include magnetic fields as 
well as shear from galactic rotation. Magnetic fields may substantially 
thicken swept-up shells (Ferri\`{e}re et al. 1991) and provide 
an additional pressure, which may the dominant component of the gas pressure 
at low temperatures (Heiles \& Troland 2005; Avillez \& Breitschwerdt 2004b). 
A large-scale shear due to differential rotation, when combined with weak 
magnetic fields, gives rise to the magnetorotational instability, which may 
dominate the turbulent velocity dispersion in the outer part of the disk (Sellwood \& 
Balbus 1999; Dziourkevitch et al. 2004; Piontek \& Ostriker 2005). 

Another possibly important yet ignored physical process is thermal conduction. 
Koyama \& Inutsuka (2004) argued for the importance of including 
thermal conduction by running two models with and without it, which showed 
discrepant global evolutions of quantities such as kinetic energy and total 
number of dense clouds. They further argued that, to properly include thermal 
conduction, the Field length $\lambda_F$ must be resoved. This is 
not feasible in our models, where the spatial resolution $\delta x \gg \lambda_F$. 
However, this crucial role of thermal conduction on cloud formation (via the action 
of thermal instability) may not carry over to the regime where 
the medium is externally driven. Based on 2D simulations that included 
stellar energy injection, V\'{a}zquez-Semadeni et al. (2000) suggested
that stellar-like forcing erases the signature of the instability, although their 
models also did not resolve $\lambda_F$. Avillez 
\& Mac Low (2002) and Avillez \& Breitschwerdt (2004a) also argued that 
turbulent diffusion may be more important than thermal conduction for mixing 
gas of different temperatures. 

\section{Results}
\label{results}

\subsection{Vertical Direction}
\label{vertical}

For the first few tens of megayears of the simulation, the heating from supernovae 
is insufficient to counter-balance the gravity. As a result, the disk initially 
cools and all gas collapses 
towards the midplane. After the collapse, shock waves bounce back and propagate 
away from the midplane in both directions, heating the gas at high 
altitudes. Almost identical behaviors were observed by Avillez (2000). As more 
supernovae explode, turbulent and 
thermal pressures build up in the disk. The total kinetic energy in the 
box exceeds the thermal energy at $t \approx$ 15 Myr and thereafter. By 
$t\approx 30$ Myr, enough supernovae have exploded that the blast waves from the 
multiple explosions permeate the disk and their interactions regulate the 
global gas structure. The average gas temperature rises steadily over this 
period. After $t \approx 60$ Myr, the system reaches a statistical steady state,
until the end of our simulation at $t \approx 80$ Myr.
The entire run took $\sim$8$\times$10$^4$ CPU hours on the TeraScale Computing 
System at the Pittsburgh Supercomputing Center.

\begin{figure*}
\hspace{2mm} 
\plotone{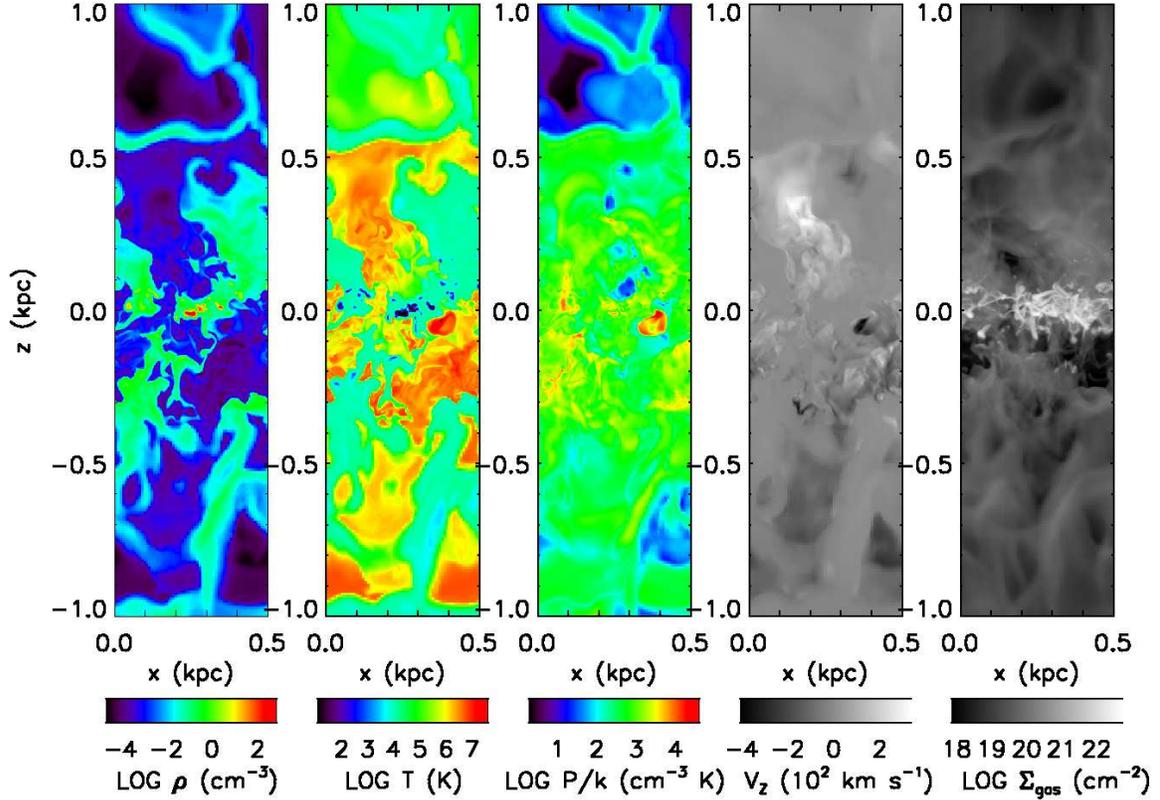}
\caption{Vertical slices of (a) density, (b) temperature, (c) pressure, (d) $z$ 
component of the velocity, and (e) column density of gas ($\Sigma_y = \int \rho \, dy$) 
at $t=79.3$ Myr. The large shells seen at $z \gtrsim 0.4$ kpc are due to two high-altitude 
Type I supernovae that exploded 3.9 and 1.5 Myr ago. 
Note the presence of round cloudlets in (b) and (e) about 100-250 pc 
away from the Galactic plane. They are fragmented (super)shells and falling 
towards the midplane at several tens of km s$^{-1}$. Filaments of neutral gas are 
found up to $\sim$2 kpc away from the midplane. Note that only the inner 2 kpc of 
our 10 kpc vertical grid is shown. 
\label{figver}}
\end{figure*}

Figure \ref{figver}(a, b, c) shows typical density, temperature, and pressure variations in 
the vertical direction, taken at $t=79.3$ Myr. A clumpy layer of cold dense 
clouds with a thickness of $\sim$200 pc sits in the midplane. 
Above and below this layer, a galactic 
fountain (Shapiro \& Field 1976) is set up, as observed in the Milky Way, 
enabling disk-halo interactions. In our model employing the Galactic supernova 
rate, the time-averaged mean mass flux through any horizontal surface approaches 
zero, consistent with the negligible fraction of gas escaping the computation box 
(see below). 

The low density gas, which 
permeates the bulk of the volume, moves away from the midplane with typical 
$v_z \approx 10^2$ km s$^{-1}$. Some expanding superbubbles are launched with 
$z$-velocities reaching 300--400 km s$^{-1}$ at the bases. These large bubbles 
subsequently break out of the disk and directly dispose their thermal energy 
content to the halo. As the dense gas (predominantly fragments of supershells) 
rises, its temperature drops because of the decreasing diffuse heating rate as 
well as radiative cooling of the gas. As a result, neutral gas is present 
up to $\sim$2 kpc above and below the midplane. 

\begin{figure}
\hspace{-1mm} 
\includegraphics[scale=0.5]{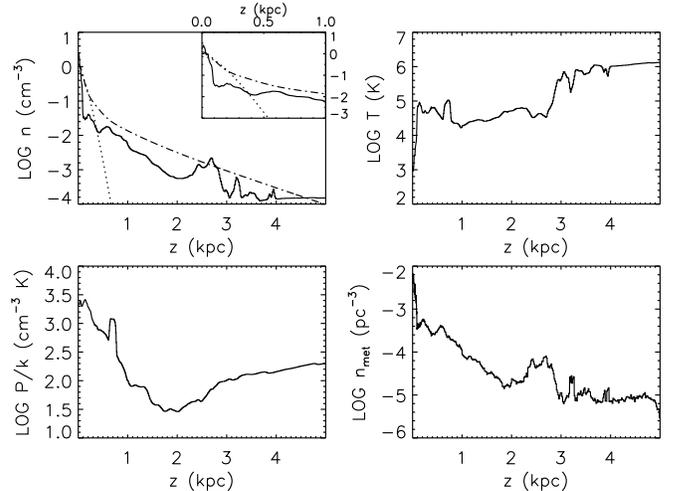}
\caption{Vertical profiles of (a) gas density, (b) temperature, (c) pressure, and 
(d) metal density at $t = 79.3$ Myr averaged over $x$-$y$ planes at constant heights. 
In (a), the dotted line represents our initial isothermal density profile, while the 
dot-dashed line shows the observed vertical profile of gas, which is the sum of 
molecular, neutral, and ionized gases. The inset displays an expanded view of the region 
near the midplane, $|z| \le 1$ kpc.
\label{figprofile}}
\end{figure}

\begin{figure*}
\hspace{-1mm} 
\includegraphics[scale=1.1]{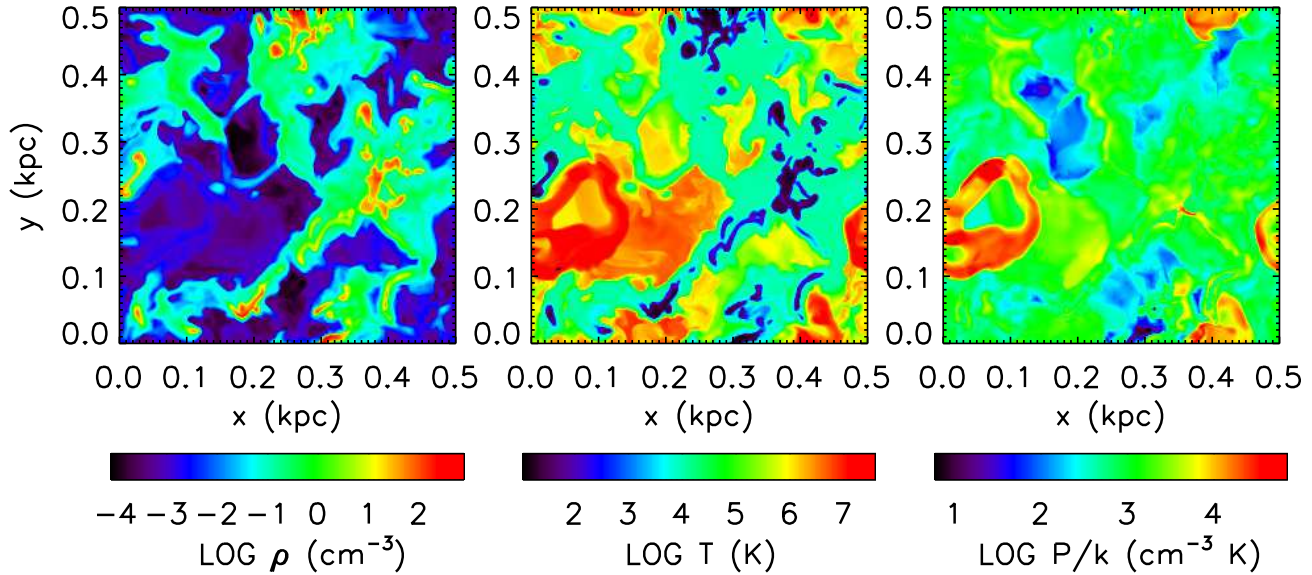}
\caption{Cuts through the midplane ($z=0$) showing distributions of (a) density, 
(b) temperature, and (c) pressure at $t=79.3$ Myr. Density 
and temperature both vary by about seven orders of magnitude. Because of an 
approximate inverse relationship between the two quantities, thermal pressure 
changes by 2--3 orders of magnitude. This much variation in pressure is 
inconsistent with ISM models based on full pressure equilibrium between phases. 
Correlated supernovae are responsible for producing most of the hot gas in the 
midplane. 
\label{figmidplane}}
\end{figure*}

Some of that gas later condenses and returns to the disk. This process occurs 
in several forms. First, cold ($T \approx 10^2$ K in the core) and dense 
($n \approx 1$ cm$^{-3}$) cloudlets can be seen 100--250 pc away from the midplane 
in Figure \ref{figver}(b, e). These 
cloudlets fall towards the midplane at velocities of a few to several tens 
of km s$^{-1}$. They typically occupy a small volume $\sim$(10 pc)$^3$, 
contain a mass 10-10$^3$ M$_{\odot}$, and are structured as a cold dense core plus 
a lower density tail, embedded in warmer gas. 
In addition, filaments of neutral clouds are found at larger heights ($|z| \approx 
0.25$--2.0 kpc). With temperatures of order a few thousand degrees, and average densities 
$\bar{n} \approx 0.3$ cm$^{-3}$, several orders of magnitude higher than their 
surroundings, these elongated filaments are estimated to contain 10$^3$--10$^5$ 
M$_{\odot}$ of neutral gas. Some extend beyond several hundred pc in length. 
Their sizes generally increase with height, but this may be an artificial 
effect caused by change in spatial resolution. It is tempting to associate the 
objects within $|z| \lesssim 1$ kpc with intermediate velocity clouds (Wakker 
2001). Neutral clouds at even larger heights ($|z| \gtrsim 1$ kpc) may be identified 
as the tangent point clouds found throughout the inner Galaxy (Lockman 2002). These clouds 
are thought to contain a large fraction of the neutral gas in the Galactic halo. 
Given the crude approximations we make with regard to heating and cooling for the 
cold dense gas, we consider this a reasonably successful test of the model. 

At high altitudes ($z \gtrsim 1$ kpc), extremely hot diffuse gas in 
the halo has 
very long cooling times $t_{cool}(n)=51/n_{-3}$ Myr, where $n_{-3} \equiv 
n/(10^{-3} \cm^{-3})$, if we take the cooling rate at $T=10^6\K$. 
Avillez \& Breitschwerdt (2004a) ran a similar model for a much longer period 
(400 Myr) and showed that after $t \approx 70$ Myr the statistical properties related 
to the vertical distribution of gas do not change appreciably.

Figure \ref{figprofile} shows how the gas density, temperature, pressure, and 
metal particle density averaged over the $x$-$y$ plane vary in the vertical 
direction. For comparison, we show in the dot-dashed line 
the observed vertical profile of gas, which is the sum of three components: 
molecular (Clemens et al. 1988), neutral (Dickey \& Lockman 1990), and 
ionized (Reynolds 1991) gases. The average density near the galactic midplane in our model 
is higher than the observed value by 2--3, while the density is somewhat underpredicted at 
a few disk scaleheights (0.1 kpc $\lesssim |z| \lesssim$ 0.5 kpc). Uncertainties in 
observations are large; see, e.g., Figure 10 of Dickey \& Lockman (1990). 
The discrepancy, if real, 
implies that supernova driving alone cannot quite provide the necessary support in the 
vertical direction to explain the observed distribution of gas. Additional components 
of pressure, e.g. from the magnetic field and cosmic rays, are expected to contribute 
significantly (Boulares \& Cox 1990). 
Compared to our initial isothermal density profile (dotted) in equation \ref{isot}, 
a larger amount of gas is present at high altitudes ($|z| \gtrsim 0.5$ kpc) 
due to the presence of a galactic fountain (Fig. \ref{figver}). 

Since outflow boundary conditions are used at the top and bottom surfaces, 
once the gas escapes with $v_z>0$, it never returns to the box. 
For the Galactic supernova rate, only a negligible fraction of the gas 
($\sim$0.1\%) escapes the computation box after 79.3 Myr of evolution. 
None of the $\sim$3$\times$10$^5$ metal particles created during the 
simulation escapes the grid boundaries at $z = \pm5$ kpc. 
The vertical distribution of metal particles peaks at the disk midplane, just 
as that of the gas density does. However, the relative concentration of the 
metal particles is substantially less pronounced, i.e., an enhancement factor 
$\sim$10$^{3.0}$ instead of $\sim$10$^{4.3}$ for the gas density, implying that 
metals are mostly associated with the warm or the hot gas phases that have 
larger scale heights than the cold gas.

For comparison, we ran a model with eight times the Galactic supernova  
rate and four times the gas column density, scaled so that the Kennicutt 
relation is approximately satisfied for the computation box. 
Galactic outflows of mass and metals are observed in this model. 
We will discuss the vertical structure of the gas in models with higher 
supernova rates in Paper II.

\subsection{In the Midplane}
\label{midplane}

Figure \ref{figmidplane} displays distributions of density, temperature, and 
pressure on a horizontal slice through the midplane ($z=0$) at $t=79.3$ Myr. 
As the standard blast wave theory (Ostriker \& McKee 1988) predicts, 
when the age of a supernova remnant approaches the cooling time of the 
shock-heated gas, the outer part of the bubble undergoes thermal instability 
and develops a dense spherical shell. These shells collide with one another to 
form clouds that are filamentary in shape. The clouds have characteristic lengths 
of several tens to hundred parsecs. Note that these cold 
dense clouds form even in the absence of self-gravity, as previous numerical 
simulations have found (Rosen et al. 1993; 
Rosen \& Bregman 1995; Korpi et al. 1999; Avillez 2000). They are formed 
due to the collective effect of thermal instability and supersonic turbulence. 
These clouds have temperature lower than $50 \K$, H$_2$ formation times of 
$\sim$1 Myr (Hollenbach \& McKee 1979), and estimated masses ranging 
from $\sim$10$^3 M_{\odot}$ to $\sim$5$\times$10$^5 M_{\odot}$. Hence, they 
can be identified as molecular clouds or giant molecular complexes observed 
in the ISM. Although in reality many of these clouds contain 
small scale sub-structures, such details are unresolved in our model. 
An analysis of the velocity field reveals that turbulent flows within the clouds 
are trans-sonic or supersonic on the cloud scale (see Fig. \ref{fig_grav}c; Paper II).

Comparing Figures \ref{figmidplane}(a) and (b) reveals an approximate 
inverse relationship between density and temperature. The highest 
gas density reaches $\sim$10$^3 \cm^{-3}$, and the temperature ranges 
from $10 \K$ to $\sim$10$^7 \K$. Both of these quantities vary by more than 
six orders of magnitude in the midplane. If they had an exactly inverse 
relationship, the thermal pressure would have been constant. However, as shown in 
the pressure map (c), there are regions that clearly depart from pressure equilibrium. 
High pressure regions are usually associated with interiors of newly 
formed supernovae, shock-compressed regions, or dense cloud cores, whereas 
low pressure regions are often interiors of old 
supernova remnants. Overall, thermal pressure varies by 2 to 3 orders 
of magnitude, confirming the results of previous works (Padoan 
et al. 1997b; Passot \& V\'{a}zquez-Semadeni 1998; Mac Low et al. 2005). This 
much variation in pressure is inconsistent with ISM models based purely on 
pressure equilibrium between phases induced by thermal instability 
(Field et al. 1969). 

\begin{figure}
\hspace{-22mm} 
\includegraphics[scale=0.79]{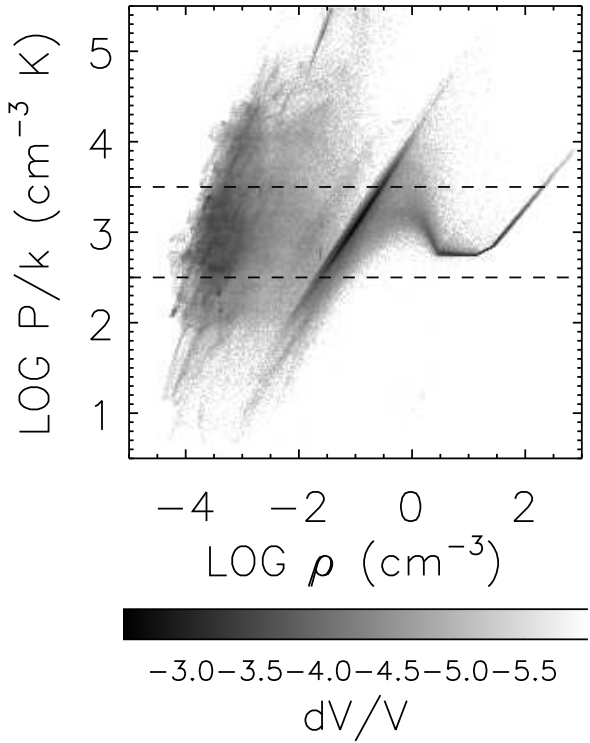}
\caption{Phase diagram showing the distribution of thermal pressure and density in the 
galactic midplane computed from eight time slices spanning 7.0 Myr of time interval 
(from $72.3$ to $79.3$ Myr). The grey scale shows the logarithm of the occupied gas 
volume. A considerable amount of gas is found in the thermally unstable regime ($200 
\le T \lesssim 1.0 \times 10^4 \K$). The straight lines in the upper left quadrant of the 
diagram are caused by the adiabatic expansion ($P \propto \rho^{\gamma}$) of young supernova 
remnants. Hence, the lines have a slope equal to the adiabatic index of gas $\gamma = 5/3$.
Interiors of young supernova remnants are associated with low 
densities and high pressures ($P/k \gtrsim 10^4$ cm$^{-3}$ K). Thermal pressures 
in dense clouds reach values about an order of magnitude higher than the mean pressure 
in the general ISM, even in the absence of self-gravity. 
\label{fig_phase_diag}}
\end{figure}

For a detailed look at the pressure distribution, we display in Figure 
\ref{fig_phase_diag} a scatter plot of pressure vs. density (a phase diagram). 
Two stable branches of the thermal equilibrium curve ($T$$\approx$10$^4$ K 
and $10 \K \le T \le 40 \K$) are densely populated. However, other parts of the 
diagram are occupied by a dynamically determined continuum of densities and 
temperatures, rather than several discrete phases. 
This is because lower density gas simply does not 
have enough time to cool back down to its equilibrium temperature 
before it gets shock-heated again. If we assume that a supernova 
shocks, on average, an area $\sim$(100 pc)$^2$ in the midplane, for the 
Galactic supernova rate, the mean time between shock passages will be 
$\sim$1 Myr. This is much longer than the cooling time, $\sim$50 Myr, 
for the gas having $n=10^{-3} \cm^{-3}$ and $T=10^6 \K$. A new physical 
description is required for this explosion-dominated medium (Mac Low \& 
Klessen 2004). 

Remarkably, $\sim$70\% of the points lie in a 
narrow range of pressure ($P/k$) between 10$^{2.5}$ and 10$^{3.5}$ cm$^{-3} \K$. 
This implies that the low to intermediate density ($n \lesssim 10^2$ 
cm$^{-3}$) ISM on several hundred pc scales should be described by a 
nearly isobaric equation of state, instead of an isothermal one, as 
typically assumed in many recent works on galaxy formation. 
The near-isobaric character of the medium is reinforced by the distribution 
of turbulent pressure, as detailed in Paper II. 


Physical quantities in our model can be compared directly with existing observations. 
In particular, we focus on several global quantities measured via H I absorption line 
studies (Heiles 2001; Heiles \& Troland 2003). They found: (1) $>$48\% of the 
WNM by mass lies in thermally unstable temperature range; 
(2) $\sim$60\% of all neutral hydrogen (again, by mass) is in the WNM; and (3) 
The warm gas occupies $\sim$50\% of the volume in the Galactic plane. 

\begin{figure}
\hspace{-15mm} 
\includegraphics[scale=0.59]{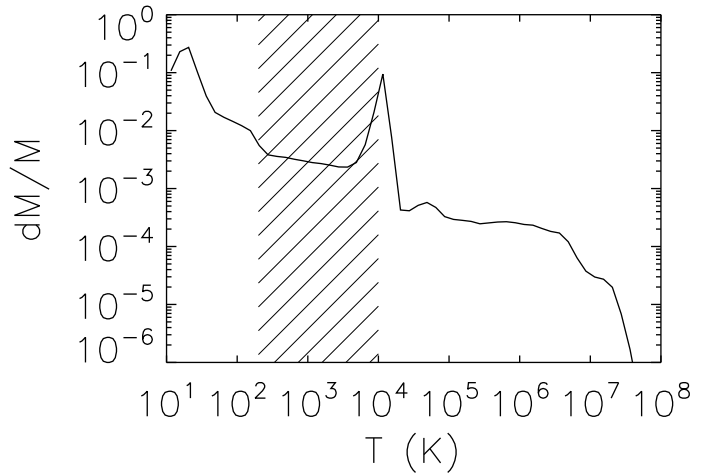}
\caption{Fractions of mass contained in logarithmically spaced temperature bins 
for gas within 500 pc of the midplane. 
The hatched region indicates the thermally unstable regime for our cooling curve. 
Note that the temperature range is more extended than in Heiles (2001), where it 
ranged from 500 to 5,000 K. Some fraction of the cold gas ($X_m$) having 
temperatures below $\sim$25 K is expected to be molecular hydrogen. 
\label{figmfrac}}
\end{figure}

\begin{figure*}
\begin{center}
\includegraphics[scale=0.85]{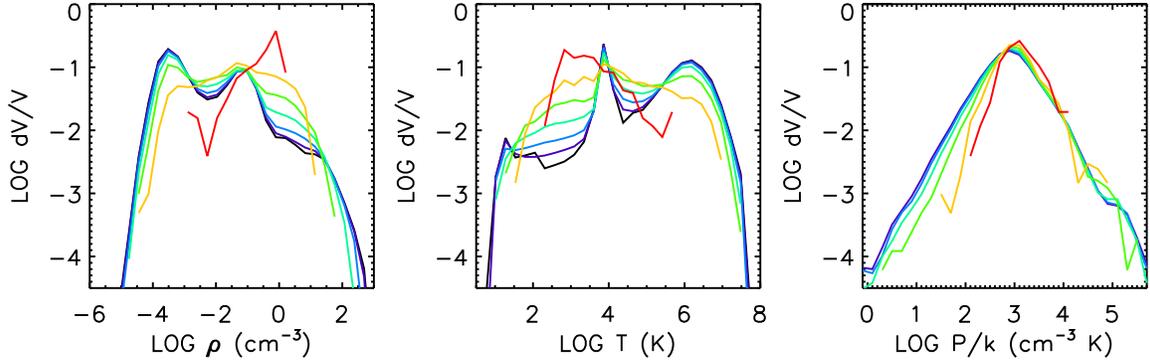}
\end{center}
\caption{Probability density functions for (a) gas density, (b) temperature, and (c) 
pressure of gas near the midplane ($|z| \le 125$ pc), taken from eight snapshots spanning 
7.0 Myr (from 72.3 to 79.3 Myr). The sizes of subboxes (kernels) used to construct the 
PDFs increase by factors of 2 from black (1.95 pc) to red (125 pc). 
The temperature PDF (b) demonstrates the extent to which the classical three-phase medium 
description is valid. 
The pressure PDF (c) shows that there is as much volume below the average pressure as 
there is above it. 
\label{figpdf}} 
\end{figure*}

The distribution of mass in terms of gas temperature is displayed in Figure 
\ref{figmfrac}. Based on the shape of our cooling curve, $T = 200$ K separates 
the CNM from WNM. The WNM ranges from 200 K to 1.7$\times$10$^4$ 
K. Some fraction of the cold gas is expected to be molecular hydrogen. 
According to Heiles (2001), most of the mass in the CNM is contained in gas with 
temperatures between 25 K and 75 K. It is then reasonable to assume a large fraction of 
colder gas ($T < 25$ K) to be molecular hydrogen. We denote this mass fraction by $X_m$, and argue 
that $X_m$ is close to unity. 
Adopting these definitions, we find: (1) 67\% of the WNM by mass is 
in the thermally unstable regime; (2) 53\% of all H I is contained in the WNM for $X_m$=1 
(43\% for $X_m$=0.9); and 
(3) The warm gas occupies 43--52\% (49\% on average) of the volume in the midplane. 
Overall, the results are broadly consistent with the observations despite our neglect of 
magnetic fields and self-gravity.

Typical hot gas filling factors in the midplane of our model $f_h \approx$ 41--51\% 
($\langle f_h \rangle = 44$\%; see Fig. \ref{figmidplane}b). These values of $f_h$ 
are lower than the estimate by McKee \& Ostriker (1977) ($f_h \approx$ 70\%). 
They ignored supernova 
clustering and the vertical density stratification, which are expected to reduce 
$f_h$. For example, models that include superbubble evolution predict 
$f_h \approx$ 17--20\% near the solar circle (Heiles 1990; Ferri\`{e}re 1998). 
The interstellar magnetic pressure will further decrease the volume occupied by 
hot gas by resisting the expansion of bubbles 
(McKee 1990; Ferri\`{e}re et al. 1991; Slavin \& Cox 1993). 
Existing observations of the hot interstellar gas are too limited to constrain 
$f_h$, but in external galaxies it is thought that $f_h \lesssim 0.5$ (Dettmar 
1992; Brinks \& Bajaja 1986).

Contrary to our result, Avillez \& Breitschwerdt (2004a) obtained low $f_h 
\approx 0.2$ at 1.25 pc resolution. Their model is discrepant from ours in 
at least the following three aspects, each of which contributes to lower 
$f_h$: (1) Their background heating rate was higher by a factor of $\sim$18; 
(2) Their supernova rate was lower by $\sim$1.7; and (3) They allowed 
Type II supernovae to explode only in cold ($T \le 100 \K$) and dense ($n \ge 
10 \cm^{-3}$) regions, while we used random locations. 
Higher background heating rates effectively shift the thermal equilibrium 
curve (see Fig. \ref{fig_phase_diag}) diagonally---upward and to the right. 
As a result, the stable branch of the curve at $T \le 40 \K$ mostly lies 
outside the pressure range occupied by the bulk of the gas. Since the mean 
thermal pressure in the midplane does not change when $\Gamma$ is increased, 
this leads to less gas in the cold phase and more gas in the warm phase. 
We experimented by running two models with cooling curves determined using ionization 
fraction of 10$^{-1}$, 
as adopted by Avillez \& Breitschwerdt (2004a). When we applied 4.5 times the diffuse 
heating rate of our fiducial model, $f_h \approx$ 50--60\%. We then restarted 
the model at $t=68$ Myr, and increased the diffuse heating rate to 18 
times the fiducial value. After only 5 Myr of evolution, the volume filling fraction 
of the hot gas dropped to 35\%. 
Instead, the warm gas occupied 50\% of the midplane area. However, we believe that 
this elevated level of diffuse heating is unphysical. 
Using density and temperature thresholds for supernova locations 
lowers $f_h$ because (1) The supernovae then destroy massive clouds, returning 
the gas to the intercloud medium; and (2) Supernovae that occur in dense regions 
occupy smaller volumes when integrated over their lifetimes (Clarke \& Oey 2002). 
The results in Korpi et al. (1999) can be interpreted using similar arguments.


\subsection{Density Fluctuations}
\label{density}
\subsubsection{Probability Density Function}

There are numerous ways to characterize density distributions of gas. 
The simplest and the easiest to interpret is the probability density 
function (PDF) for the gas density. It is displayed in Figure \ref{figpdf}(a). 
We use various subbox sizes over which density PDFs are computed. Purple 
corresponds to the smallest cubic subboxes with 3.91 pc on a side, and red 
to the largest subboxes with 125 pc on a side. 

We find that most of the simulation volume is occupied by low-density 
gas due to supernova feedback, in accord with Slyz et al. (2005), 
who simulated turbulent interstellar medium models in a non-stratified 
(1.28 kpc)$^3$ box with periodic boundary conditions and $\ge$10 pc resolution. 
Stellar feedback and/or self-gravity of gas were included in some of their models. 
In terms of the density PDF, they found that (1) The models without stellar 
feedback showed log-normal PDFs with a single peak; (2) A power-law tail developed 
in the high-density end when self-gravity was included; and (3) The PDF became markedly 
bimodal when stellar feedback was included, regardless of whether self-gravity was 
included or not. 

Although our PDF is not bimodal, there is a hint of a broad peak near $n = 10 
\cm^{-3}$ for the (thermally stable) cold high-density gas.
It is controversial if this part of the PDF can be approximated by a log-normal 
function. Using 2D simulations including heating, cooling, and self-gravity, 
Wada \& Norman (2001) claim the high-density end of their density PDFs is 
well-fitted by a log-normal function (see their Figure 16). In contrast, 
several numerical experiments that included either self-gravity of gas or 
non-isothermal equation of state ($\gamma \ne 1$) reported that high-density 
tails develop in non-isothermal cases (Scalo et al. 1998; Li et al. 
2003). Although a log-normal density PDF is a natural outcome for isothermal gas 
(Passot \& V\'{a}zquez-Semadeni 1998), support for such a PDF for non-isothermal 
cases remains weak. 

The one-point density PDF does not contain information on how dense regions 
or voids are connected in space. For this reason, the cloud mass spectrum cannot 
be derived from the density PDF alone, as Scalo et al. (1998) point out. 
To supplement the density PDF, we attempt two methods of analysis. First, Figure 
\ref{figpdf}(a) shows how the density PDF changes as the subbox size 
increases. That the PDFs show markedly different shapes attests to the importance 
of specifying the smoothing scale with which the density PDF is measured. More 
significantly, the Jeans mass $M_J$ and the density threshold for gravitational 
collapse $\rho_{th}$ also change as a function of scale (see Fig. \ref{fig_grav}).
Despite the fact that we do not fully understand what determines the shapes of 
the PDFs as the subbox size increases, this clearly has implications for 
gravitational collapse, as we explore in \S\ \ref{clouds}. 

\subsubsection{Power Spectrum}

The other method to characterize density fluctuations is to use second-order 
statistics such as the auto-correlation function or its Fourier transform (FT), 
the power spectrum. Our computation box is periodic in 
the $x$ and $y$ directions but non-periodic in $z$, while FFTs assume 
periodicity in all directions. Hence, before taking FFTs, we apply 
to the $z$-components of the variables the Hanning window defined by 
$w(z) = (1/2)(1+\cos(2\pi z/L_z))$ where $L_z$ is the vertical extent of 
the box used for the FFT (0.5 kpc) and $-L_z/2 \leq z < L_z/2$. 

While the velocity power spectrum in an incompressible medium $|\bov(\bx)|^2$ 
has the clear physical meaning of specific kinetic energy (see the discussion 
of Parseval's theorem in \S\ \ref{driving}), $\rho(\bx)^2$ does not. 
If the density PDF turns out to be lognormal and if $s \equiv \log \rho$ can 
be taken as a Gaussian random variable, a power spectrum of $s$ can 
completely specify the density distribution. Even though this is 
not the case (see Fig. \ref{figpdf}a), the density power spectrum is 
nevertheless useful for comparison with previous numerical models. 

\begin{figure}
\hspace{-1mm} 
\includegraphics[scale=0.5]{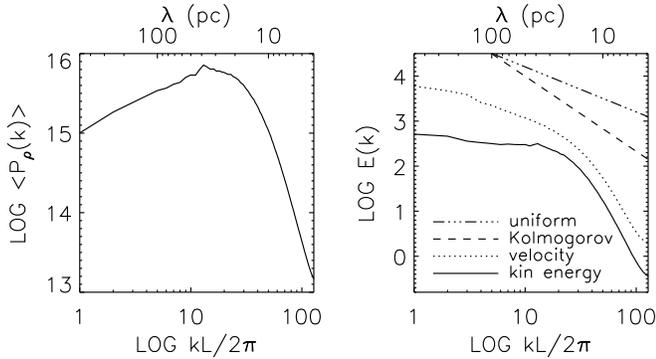}
\caption{(a) Angle-averaged density power spectrum, displaying a wide peak 
around $kL/2\pi \approx 20$. The box size $L = 0.5$ kpc. The power falls off at large 
wavenumbers (small wavelengths) due to numerical diffusion. 
(b) Kinetic energy spectrum (solid) and angle-averaged velocity power spectrum (dotted). 
The kinetic energy is distributed over a wide range of wavenumbers. There is no single 
effective driving scale, unlike in Kolmogorov's idealized picture of incompressible turbulence. 
We find that 90\% of the total kinetic energy is contained in wavelengths $\lambda \le 
190$ pc. Because of the highly intermittent density structure, the velocity power spectrum 
is not parallel to the kinetic energy spectrum, especially at large scales. 
To guide the eye, two straight lines are plotted: the Kolmogorov energy spectrum (dashed) 
and the spectrum containing an equal amount of energy per decade (dot-dashed). 
\label{figps}}
\end{figure}

The density power 
spectrum of an explosion-driven, strongly compressible medium is displayed in 
Figure \ref{figps}(a). Its shape contrasts drastically with its 
counterpart in a weakly compressible medium, where density fluctuations 
behave as a tracer field and 
possess a Kolmogorov spectrum (Lithwick \& Goldreich 2001). The spectrum 
peaks near the smallest scale at which kinetic motions can be well-resolved 
$\lambda_{\rho} \approx 20$ pc. 
This wavelength is slightly smaller than the most energy-containing 
scales $\lambda_{E} \approx 20$--40 pc (Fig. \ref{figps}b). 
Since $\lambda_{\rho}$ is within the range affected by numerical diffusion 
(\S\ \ref{driving}), this conclusion is somewhat uncertain. 
We do note that our value of $\lambda_{\rho}$ is comparable 
to the correlation length of 28 pc measured from two-dimensional 
autocorrelation functions of the molecular hydrogen column 
density map in the Taurus molecular complex (Kleiner \& Dickman 1984). 

It is simple to predict the density power spectrum of a medium dominated 
by shocks, because a shock is essentially a step function in density and 
velocity fields. Step functions are generally associated with power spectra 
having $P(k) \propto k^{-2}$ (Passot et al. 1988). However, 
neither observations nor numerical simulations support this relation for 
density. Padoan et al. (1997a) showed 
that the density power spectrum has a power-law form $P(k) \propto k^{-\alpha}$
with $\alpha = 2.6 \pm 0.5$, similar to values determined from observations 
(Stutzki et al. 1998). However, comparison with observations is not straightforward, 
since only the projected (column) density, not the physical density, is available 
from observations. The steep spectral 
slope at larger wavenumbers may have been caused by the nature of PPM; see 
\S\ \ref{driving}. Unlike the spectrum in Padoan et al. (1997a) who used a periodic 
isothermal box, our density spectrum turns over at 
$kL/2\pi \approx 10$ and falls off at small wavenumbers (large wavelengths).

Finally, we add a cautionary note. 
Generally, one should be careful when computing power spectra of 
physical quantities from an AMR code. This is because power at small scales 
can be easily underestimated in relatively under-refined regions. To avoid 
such mistake, the choice of 
refinement and derefinement criteria is crucial (Kritsuk et al. 
2004). Although our simulations are performed with an AMR code, 
the region near the midplane, particularly the cube with volume (500 pc)$^3$ 
within $|z| \le 250$ pc, 
ends up refined uniformly to the maximum refinement level after a few 
tens of Myr of evolution due to extreme density inhomogeneities and 
turbulent motions caused by the explosions. (For this reason, in fact, 
we effectively turn off the AMR machinery at late times.) Thus, at late 
times, we may compute power spectra without misrepresenting them.

\subsection{Energy Fluctuations: Driving Scale}
\label{driving}

How is the kinetic energy distributed among various length scales in 
our turbulent medium stirred by supernova explosions? This question is crucial 
for star formation. Depending on the slope of the forcing function $k^{-\alpha}$, 
the statistics of turbulent fluctuations and clumps change (Bonazzola et al. 1987; 
Biferale et al. 2004).

Previous simulations of uniformly driven (as opposed to 
explosion-driven) turbulence with variable driving wavelengths 
have shown that the occurrence and efficiency of local collapse 
into dense cores---and, presumably, stars---decreases as the 
driving wavelength decreases (KHM00). 
Despite its relevance to cloud formation, as far as we know, the driving 
scale has never been measured in 
numerical simulations of ISM driven by point explosions. 
Observations of molecular clouds show that power-law scaling extends up to the largest 
observed clouds (Ossenkopf \& Mac Low 2002), and suggest 
that the turbulence is driven by large-scale, external sources.
However, gas density and velocity---two physical quantities relevant to the 
kinetic energy density---can only be disentangled by observations under 
certain conditions (Brunt \& Mac Low 2004; Lazarian \& Esquivel 2003).

In an incompressible or weakly compressible medium, gas density is 
nearly uniform and thus the energy spectrum is simply the angle average of the 
velocity power spectrum: $E_{inc}(k) = 4\pi k^2 \langle P_v(\bk)\rangle _{\Omega}$ 
where $k = |\bk|$ and $P_v(\bk)$ is the three-dimensional Fourier transform of 
each component of $\bov$, squared and then summed.
The velocity power spectra of compressible interstellar gas have been studied 
in 2D (Dahlburg et al. 1990; V\'{a}zquez-Semadeni et al. 1996; Wada \& Norman 
2001) and in 3D (e.g., Kritsuk et al. 2004). However, none of the 
previous simulations performed in 3D dealt explicitly with power spectra of 
the medium driven by discrete physical space forcing. 

In a strongly compressible medium with high Mach numbers, the gas density 
$\rho$ is highly intermittent and can vary by several 
orders of magnitude. To measure how much kinetic energy is 
contained near a given wavenumber $k$, we should compute the power spectrum 
of $\sqrt{\rho} \bov$ instead of $\bov$, since according to Parseval's theorem: 
\begin{equation}
\int_{-L/2}^{L/2} \left[ f(x) \right]^2 \, dx = \int_0^{\infty} \left[ 
|\hat{f}(k)| \right]^2 \, dk \; , 
\end{equation}
where $f(x)$ and $\hat{f}(k)$ are Fourier transform pairs, and $L$ is the size of 
the periodic box. If we substitute $v_i(\bx)$ for $f(x)$ (where $i=x, \, y, \, z$), 
we obtain a relation for the incompressible case, which justifies the use of 
the velocity power spectrum for the kinetic energy spectrum. If we now 
substitute $\sqrt{\rho} v_i$ in place of $f(x)$, we obtain the corresponding relation 
for a compressible medium, which relates the kinetic energy density ($\rho v_i^2$) to the 
Fourier transform of $\sqrt{\rho} v_i$. As we demonstrate below, the power spectrum of 
$\sqrt{\rho} \bov$ shows a markedly different shape from that of $\bov$. This point was 
briefly noted in the discussion of Padoan \& Nordlund (1999). 

Figure \ref{figps}(b) shows the one-dimensional kinetic energy spectrum 
$E_{kin}(k) = 4\pi k^2 \langle P_w(\bk)\rangle _{\Omega}$ where 
$w \equiv \sqrt{\rho} \bov$. 
The measured slope is far shallower than the Kolmogorov spectral index of 
$-5/3$ for incompressible turbulence. In a strongly compressible, explosion-driven 
turbulence, shocks and interacting blast waves contribute to a wide 
range of driving wavelengths. The shallow spectral slope shows the presence 
of substantial power at small length scales, suggesting that SN-driven motions 
dominate the kinetic energy spectrum even at small scales 
where H II regions (Haverkorn et al. 2004) or protostellar outflows may 
also be active. Thus, turbulence in the dense medium is at least partly driven by 
supernovae. 

The angle-averaged velocity power spectrum often used in the literature is 
also plotted in Figure \ref{figps}(b) for comparison. This spectrum contains 
velocity information only and neglects the density inhomogeneity, as though the 
medium were incompressible. The remarkable difference at low wavenumbers between 
the two spectra in Figure \ref{figps}(b) suggests that 
dense regions, which more or less determine the shape of the kinetic energy 
spectrum, have a vastly different velocity structure from rarefied (intercloud) 
regions.

Although kinetic energy is distributed over a broad range of scales, 90\% 
of the total kinetic energy is contained on wavelengths $\lambda \le 190$ pc. 
This value (190 pc) is likely to decrease somewhat in higher resolution simulations 
that better resolve turbulent motions on $\lesssim 20$ pc scales. In typical 
cosmological simulations, therefore, most of the turbulent energy due to multiple 
supernova explosions is contained in scales smaller than a resolution element. 
To include the effect of hydrodynamic feedback in such simulations, it is necessary 
to use a subgrid model that represents unresolved, small-scale motions (Paper II). 

To test the robustness of our results on numerical resolution, we ran our 
fiducial model with the same box size at three different resolutions (8, 4, and 2 pc). 
The energy power spectra for the three runs 
display identical shapes in the dissipation range, but 
are shifted horizontally with respect to one another by $\sim$2. 
This occurs because, in a compressible medium, structures on scales smaller 
than $\sim$10 grid zones are affected by numerical dissipation (Porter
et al. 1992; Mac Low \& Ossenkopf 2000), determining the inertial range of the 
turbulent cascade. Finite difference algorithms directly damp kinetic motions on 
small scales and convert their kinetic energy into heat. 
It follows that velocity differences on small scales ($\lesssim10\Delta x$) 
are somewhat underestimated. However, that may not affect the overall dynamics 
since thermal pressure should dominate the turbulent pressure on such small scales. 

The power spectra tend to flatten slightly near the wavenumber of maximum dissipation. 
This distinctive shape of the spectrum in the dissipation range ($\lesssim 10\Delta x$) 
is caused by the bottleneck effect known to occur in hydro codes 
using PPM (Falkovich 1994; Porter et al. 1992; Porter et al. 1998).

\subsection{Velocity Structure Function}

\begin{figure}
\includegraphics[scale=0.53]{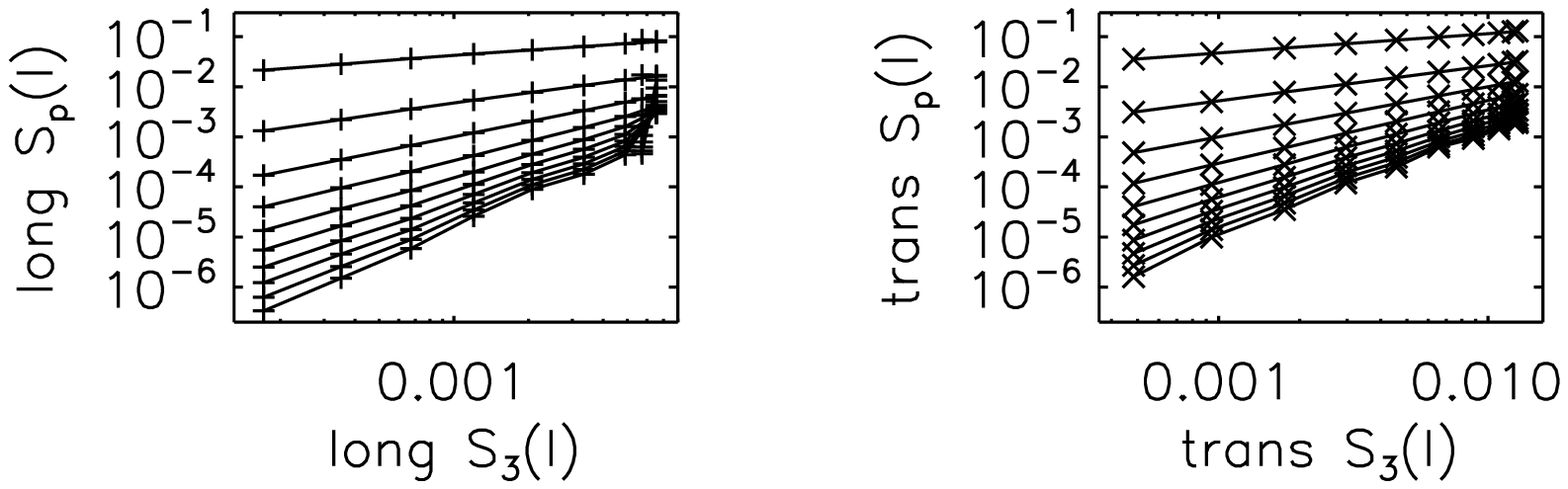}
\includegraphics[scale=0.55]{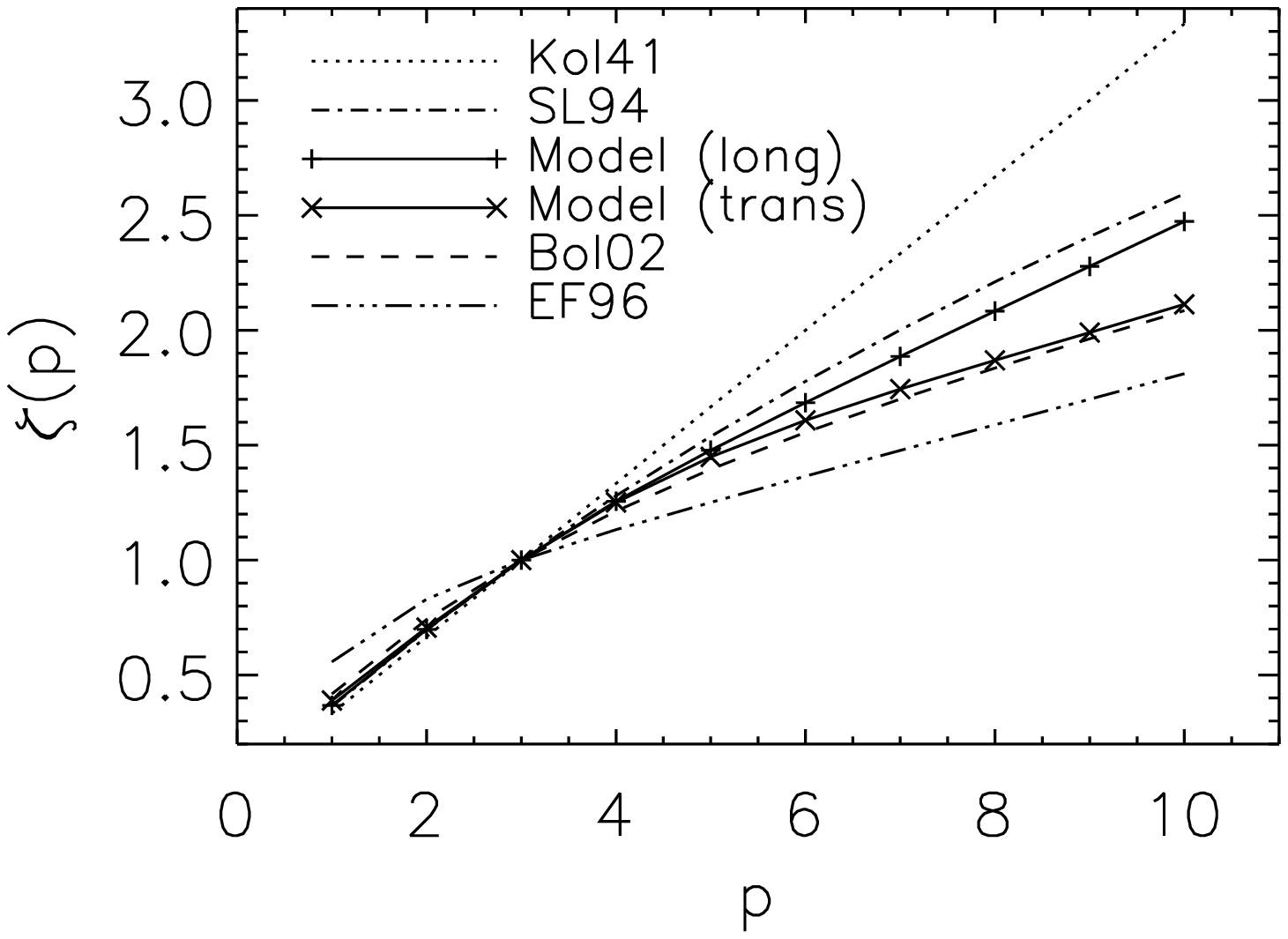}
\caption{(a) Longitudinal and transverse velocity structure functions. (b) Scaling 
exponents $\zeta(p)$. Theoretically expected scalings are shown for comparison. 
From top to bottom, they are from Kolmogorov (1941) {\it (dotted)}, She \& Leveque 
(1994) {\it (dot-dashed)}, and Boldyrev (2002) {\it (dashed)}. 
The dot-dot-dashed line is plotted assuming $D=2.3$, taken from Elmegreen 
\& Falgarone (1996)'s observation of molecular clouds. The data points are 
from, again from top to bottom: longitudinal and transverse velocity 
structure functions. 
\label{figsf}}
\end{figure}

The velocity power spectrum is steeper than the slope expected for Kolmogorov 
turbulence ($-5/3$). In addition, since the spectral slope varies depending on 
the wavenumber, it is unclear whether an inertial range exists at all. How 
can we understand this behavior? Extensive experimental and numerical studies 
have found deviations from the Kolmogorov spectrum, usually referred to as 
intermittency corrections. To quantify deviations from the Kolmogorov spectrum, 
it is useful to compute a velocity structure function defined by  
\begin{equation}
S_p(l) = \langle|\bov(\bx+\bl)-\bov(\bx)|^p\rangle \propto l^{\, \zeta(p)} \, ,
\end{equation}
where the scalar lag $l = |\bl|$. There are several phenomenological 
frameworks for compressible turbulence, each of which gives a distinct
prediction for the scaling exponents $\zeta(p)$. We apply the idea of 
extended self-similarity, which showed correct scaling behaviors for $\zeta(p)$ 
even in systems with moderate Reynolds numbers, specifically the power index 
$-(1+\zeta(2))=-1.74$ for a relatively wide range in $S_3(l)$ (Benzi et al. 
1993; Camussi \& Benzi 1996). We plot the structure function against the 
third-order structure function $S_3(l)$ in Figure \ref{figsf}(a). 
 
Predictions for $\zeta(p)$ from various analytic theories of turbulence are 
shown in Figure \ref{figsf}(b). Based on She \& Leveque (1994)'s model of 
intermittency for incompressible turbulence (see also She \& Waymire 1995), 
Boldyrev (2002) proposed an extension, noting that the most disspative 
structures in strongly compressible ISM tend to be shocks, i.e. two-dimensional 
sheets. Our structure functions for the transverse component of the 
velocity field are most consistent with the prediction from this theory.
Those for the longitudinal component diverge from straight lines at large 
$S_3(l)$ (i.e., large separations), because most of the gas in terms of volume 
travels away from the midplane. If we omit the last two data points for each 
value of $p$, we find that the longitudinal structure functions also agree 
with Boldyrev (2002)'s prediction. 
Until our work, the theory has been compared well only with isothermal gas 
driven on large scales by a random Fourier space forcing 
(Boldyrev et al. 2002; Padoan et al. 
2003). Our result demonstrates that the theory is applicable 
even to a medium driven by discrete point explosions and 
subject to nonlinear heating and cooling processes.  


\section{Gravitationally Unstable Clouds}
\label{clouds}

The connection between small-scale star formation in local regions 
of cold dense gas and large-scale, galactic star formation is unclear. The 
Kennicutt-Schmidt law is successful in predicting star-forming rates (SFRs) 
of galaxies with wildly different characteristics---from low-surface 
brightness galaxies to starbursts---implying that the SFR can be estimated 
from macroscopic properties of galaxies, and is perhaps controlled mainly 
by gravitational instability. Such a claim was made by Li et al. 
(2005), who used an isothermal equation of state for gas with a high speed 
of sound to represent the interstellar turbulent pressure. One might argue, 
though, that star formation is inherently a local phenomenon that depends 
sensitively on the physical conditions on cloud scales. 

On kiloparsec scales, much larger than the driving scales of the interstellar 
turbulence that we measured in \S\ \ref{driving}, gravity is important and the 
effect of turbulence can be 
represented as an effective turbulent pressure term (Elmegreen \& Scalo 
2004). Below those scales, at the scales resolved by our 
numerical model, supersonic turbulence controls much of the cloudy structure.

Observational and numerical studies have established that the Toomre (1964) parameter 
$Q \equiv \kappa \sigma/\pi G \Sigma_{gas}$, where $\kappa$ is the epicyclic 
frequency and $\sigma$ is the effective sound speed, controls whether or not a 
given region of a galaxy will actively form stars (Klessen 1998; Li et al. 2005). 
However, even in Toomre-unstable regions, not all the gas collapses within the 
local freefall time. Several authors have argued that only a fixed (constant) 
fraction of mass $f_M$ above some density threshold will collapse to form stars 
(Elmegreen 2002; Krumholz \& McKee 2005), based on the universality of density 
PDFs in turbulent media (Wada \& Norman 2001). Still, 
many questions remain unanswered: what determines $f_M$; is there a clear 
density threshold for star formation; and is $f_M$ the same 
(on average) for various galaxy types?

One of the aims of our work has been to study how gas dynamics and 
thermodynamics shape the ISM. 
In our model, cold dense clouds form directly in the turbulent flow. 
What fraction of these clouds are gravitationally 
unstable, given the density and velocity structures in our model? 
If gravity is decoupled from other 
cloud-forming processes such as turbulence and thermal instability, do we 
still reproduce a star-formation rate consistent with the value implicitly set 
by the input supernova rate? To answer these questions, we apply to the data a 
simple criterion for gravitational collapse (see below). 
By comparing the collapse rate with 
our input supernova rate, we can quantitatively check the validity of our 
high-resolution simulations as a model for star formation. 


Since our model does not include self-gravity of gas, we estimate the 
gravitational collapse rate (hence the SFR) by using a crude method that 
involves the modified Jeans criterion (Chandrasekhar 1951). This approach 
will yield the correct result only if (1) 
the spatial resolution of our model is sufficiently high that the gas 
dynamics is well resolved, particularly physical quantities such as mass 
density $\rho$ and velocity dispersion of gas $\sigma$ that 
are crucial for estimating the collapse rate, and 
(2) supersonic random motions determine 
most properties of star-forming molecular clouds, 
with just minor modifications from gravity (Padoan
et al. 1997a; Padoan et al. 1997b; Padoan et al. 1998). 
If self-gravity extensively alters the structure of clouds presented in 
\S\ \ref{results}, our estimate of the collapse rate will be inaccurate. 

We analyze the gas in the rectangular region near the midplane 
within $|z| \leq 125$ pc that contains the bulk ($>$90\%) of the total 
gas mass. 
We estimate the total collapse rate of gas as follows. 
First, we identify Jeans-unstable boxes at varying scales using the criterion 
$M_{box}/M_J > 1$, where 
$M_J=\bar{\rho} \lambda_J^3$, $\bar{\rho}$ is the average density in the box, 
$\lambda_J=(\pi/G \bar{\rho})^{1/2} \sigma_{tot}$, and 
$\sigma_{tot} \equiv (\bar{c_s}^2+\frac{1}{3} \bar{\sigma}^2)^{1/2}$ (Chandrasekhar 1951).
We use $\bar{c_s}^2 \equiv (1/M_{box})\int_{box} c_s^2 dM$ and $\bar{\sigma}^2 \equiv 
(1/M_{box})\int_{box} \sigma^2 dM \,$. (Paper II contains more details on how 
$\bar{c_s}$ and $\bar{\sigma}$ are computed.) 
The total velocity dispersion $\sigma_{tot}$ changes with scale, because kinetic 
energy is distributed over a wide range of scales (\S\ \ref{driving}). 
Then the total collapse rate is estimated assuming that the unstable boxes collapse 
within the local freefall time. To convert this value to the SFR, we multiply it 
by a star formation efficiency of 0.30, as it is thought that $\sim$30\% of the cloud mass that 
gravitationally collapses is converted to stars (Li et al. 2005 and references therein).

\begin{figure*}
\begin{center}
\includegraphics[scale=0.62]{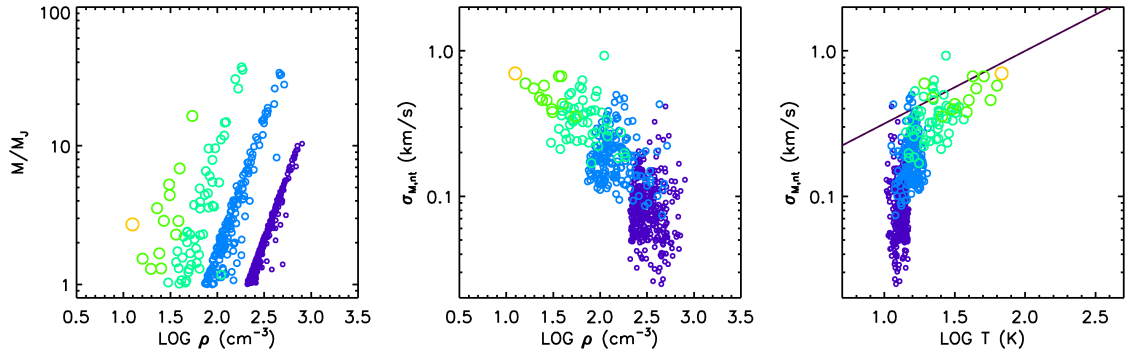}
\end{center}
\caption{(a) Mass contained in each subbox vs. average gas density of the 
subbox. There is no single density threshold for gravitational collapse. 
(b) Turbulent velocity dispersion vs. average density of gas in the subbox. 
(c) Same as (b), but plotted against the mass-weighted gas temperature. Most 
Jeans-unstable boxes on the cloud scale are associated with transonic or supersonic 
velocity dispersions. 
\label{fig_grav}}
\end{figure*}

Figure \ref{fig_grav}(a) shows that there is no single density threshold $\rho_{th}$ for 
gravitational collapse. Instead, $\rho_{th}$ depends on the scale at which collapse occurs. 
The turbulent velocity dispersions for gas with number density of 10$^2$ cm$^{-3}$ range 
from 0.1 to 1.0 km s$^{-1}$ (Fig. \ref{fig_grav}b). 
For the small (purple and blue) subboxes, the turbulent velocity dispersion is 
lower than the local sound speed (Fig. \ref{fig_grav}c), as turbulent motions on those 
scales are not well resolved in our model.

\begin{figure}
\includegraphics[scale=0.5]{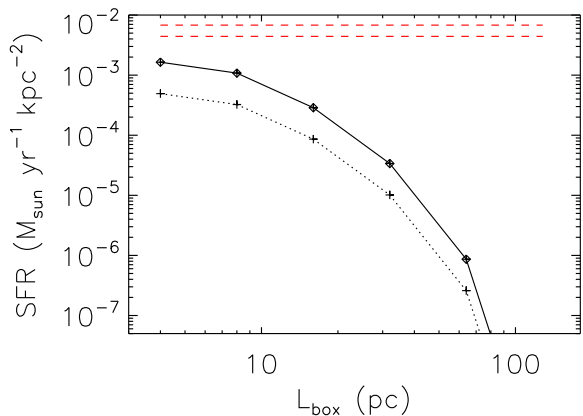}
\caption{Predicted star formation rate from the model plotted against the 
subbox sizes used. The dotted line is drawn assuming that 30\% of the mass 
in Jeans unstable regions turns into stars. The red dashed lines show the 
SFRs consistent with the assumed Galactic supernova rate, assuming 130 
or 200 M$_{\odot}$ of stellar mass is required per supernova. 
\label{fig_sfr}}
\end{figure}

The plot of the predicted SFR against the subbox sizes is 
displayed in Figure \ref{fig_sfr}. Our results 
show that, for $L_{box} \gtrsim 20$ pc where turbulent motions are resolved, 
even if all the gas in 
turbulent Jeans unstable boxes collapses and forms stars within the local 
freefall time $t_{ff}$, the resulting star formation rate remains lower than 
the value consistent with our input supernova rate: 4.4 (or 6.8) $\times 10^{-3}$ 
M$_{\odot} \yr^{-1} \kpc^{-2}$, which is computed assuming 130 (or 200) M$_{\odot}$ 
of stellar mass is required per supernova using the Salpeter IMF.
 
Turbulent compression and the structuring of the ISM by supernovae may 
set the stage for further gravitational collapse and even determine where 
it occurs. However, the discrepancy between the assumed and the predicted 
SFRs illustrates a limitation of our model. 
We interpret this as an indication
that the density PDF must be affected by self-gravity; this idea is 
supported by numerical results that the high-density end of the PDF, most 
crucial for gravitational collapse, is 
positively skewed when self-gravity is turned on (Li et al. 2003; 
Slyz et al. 2005). If this is true, computations of SFRs 
based directly on the assumption of a log-normal density PDF may be misleading. 
Uncertainties in the present analysis are missing kinetic 
energy caused by the finite resolution (\S\ \ref{driving}), which tends to 
increase predicted SFRs on small scales, and 
the absence of magnetic fields. Magnetic pressure reduces the probability 
of collapse (HMK01; V\'{a}zquez-Semadeni et al. 2005). 

We conclude that, while it may be possible to predict where gravitationally 
bound regions will occur, estimating the SFR based on our 
model leads to an underestimate, about an order of magnitude lower 
than the value consistent with the assumed supernova rate. This limitation 
of our ISM model probably stems from the key physical process that we ignore: 
self-gravity of the gas. Other models relying entirely on turbulent compression 
will likely show similar deficits.


We verified that the gas sitting at a uniform temperature ($T = 4.2 \times 10^3$ K, 
corresponding to the total (thermal plus turbulent) velocity dispersion of our medium 
$\sigma \approx 6.5$ km s$^{-1}$) in hydrostatic equilibrium with the imposed 
gravitational field is stable. In other words, no Jeans unstable boxes occur. 
However, this point cannot be used to argue 
that supernovae enhance the global gravitational collapse rate and trigger star 
formation. In fact, the opposite is true. Since the gas was initially out of thermal 
equilibrium, it would have promptly collapsed and cooled into a thin sheet. 
Most of the mass would have become gravitationally unstable, without extra 
stirring from supernovae. The net effect of supernova-driven turbulence is to 
{\it inhibit} star formation globally by {\it decreasing} the amount of mass 
unstable to gravitational collapse. 

\section{Summary}
\label{discuss}

In this work, we have constructed 3D models of the ISM including vertical 
stratification and discrete physical space forcing by supernova explosions. 
We included parametrized heating and cooling as well as both isolated and 
correlated supernovae. This work supports models that treat supernova-driven 
turbulence as a major structuring agent in the ISM. 
Our main results are as follows:

\begin{itemize}
\item Cold dense clouds naturally form as blast waves sweep the gas 
and collide. The intercloud region is filled with 
warm or hot gas having relatively low densities. 
The resulting density and velocity structures adequately 
match the observed filamentary and clumpy structure of the ISM. 

\item A galactic fountain reaching a height of several kiloparsecs forms, with 
most of the volume moving away from the midplane. Fragmented shells rain 
back down as small cold clumps, which may be identified as intermediate 
velocity clouds.

\item Global quantities in our model are broadly consistent with existing 
observations. Specifically, our results are consistent with observations of 
the mass fraction of thermally 
unstable gas and the warm gas filling factor in the Galactic plane derived 
from H I absorption lines (Heiles 2001; Heiles \& Troland 2003). 

\item The diffuse heating rate $\Gamma$ influences the global 
structure of our model. For example, in our numerical experiments, 
increasing $\Gamma$ by 4 lowered the occupation fraction of the hot gas 
in the Galactic plane by $\sim$20\%. 
This occurs because high background heating rates effectively shift the 
thermal equilibrium curve in the phase diagram so that the cold (thermally 
stable) branch lies mostly outside the range of pressure occupied by the 
bulk of the gas.

\item The vertical distribution of gas in our model deviates from the 
profile inferred from observations. Specifically, the gas is excessively 
concentrated in the disk midplane by factors of 2--3, while it is deficient 
at $0.1 \lesssim |z| \lesssim 0.5$ kpc. 
We attribute this discrepancy to neglected components of pressure 
such as those from the magnetic field and cosmic rays.

\item The density power spectrum has a broad peak close to the dissipation 
scale $\lambda \sim 20$ pc. It turns over and falls off at large scales.

\item There is no single effective driving scale; instead, we see 
energy injection over a range of scales. As a result, the kinetic energy 
power spectrum possesses a slope much flatter than Kolmogorov's 
spectrum at scales between $\sim$20 and $\sim$500 pc. 
The predominance of small scale density structure causes the kinetic energy power 
spectrum to look remarkably different from the often-used velocity power 
spectrum. About 90\% of the total kinetic energy is contained in wavelengths 
shortward of 190 pc.

\item The velocity structure functions demonstrate that the phenomenological 
theory proposed by Boldyrev (2002) is applicable even to a medium driven by 
discrete point explosions and subject to nonlinear heating and cooling processes.

\item Even if all the gas in turbulent Jeans unstable subboxes in our 
simulation collapses and forms stars in local freefall times, 
the resulting collapse rate is significantly lower than the value 
consistent with the input supernova rate. Supernova-driven turbulence inhibits, 
but does not prevent, star formation. From this, we infer that 
the density PDF must be significantly affected by self-gravity. For predicting  
physical properties of star-forming regions (such as the stellar 
initial mass function), the statistics of turbulent fluctuations 
generated by models without self-gravity can be misleading.
\end{itemize}

Applying high supernova rates to our model galaxy  
naturally leads to galactic outflows that have already been 
observed at both low (Heckman et al. 2000) and high redshifts 
(Pettini et al. 2001, 2002; Shapley et al. 2003). In the future, we 
will use these local models to measure 
the efficiency of energy transfer into superwinds out of a 
stratified disk for a given star formation rate, and thereby 
construct subgrid models for stellar energy feedback in 
cosmological simulations. 

\acknowledgments

We are grateful to M. Avillez, S. Glover, I. Goldman, C. Heiles, L. Hernquist, 
E. Jenkins, R. Klessen, Y. Li, C. McKee, J. Stone, and S. White for 
stimulating discussions. We acknowledge a helpful suggestion 
from G. Bryan regarding the importance of the thermal equilibrium curve in 
interpreting the relationship between the hot gas filling factor and 
background heating rate. We thank J. Oishi for valuable comments 
on the manuscript that improved the presentation and J. Maron for suggesting 
the use of structure functions. M. K. R. J. was supported by an AMNH Graduate 
Student Research Fellowship. M.-M. M. L. acknowledges support by NSF Career 
grant AST99-85392, and NSF grants AST03-07793, and AST03-07854. The software 
used in this work was in part developed by the DOE-supported ASCI/Alliance 
Center for Astrophysical Thermonuclear Flashes at the University of Chicago.
Computations were performed at the Pittsburgh Supercomputing Center supported 
by the NSF. 


\clearpage

\clearpage

\clearpage


\begin{thebibliography}{}
\bibitem[Abbott(1982)]{abb82} Abbott, D. C. 1982, ApJ, 263, 723
\bibitem[Armstrong et al.(1995)]{arm95} Armstrong, J. W., Rickett, B. J., \& Spangler, S. R. 1995, ApJ, 443, 209
\bibitem[Avila-Reese \& V\'{a}zquez-Semadeni(2001)]{avi01} Avila-Reese, V., \& V\'{a}zquez-Semadeni, E. 2001, ApJ, 553, 645
\bibitem[Avillez(2000)]{avi00} Avillez, M. A. 2000, MNRAS, 315, 479
\bibitem[Avillez \& Berry(2001)]{avi01} Avillez, M. A., \& Berry, D. L. 2001, MNRAS, 328, 708
\bibitem[Avillez \& Mac Low(2002)]{avi02} Avillez, M. A., \& Mac Low, M.-M. 2002, ApJ, 581, 1047
\bibitem[Avillez \& Breitschwerdt(2004a)]{avi04a} Avillez, M. A., \& Breitschwerdt, D. 2004a, A\&A, 425,899
\bibitem[Avillez \& Breitschwerdt(2004b)]{avi04b} Avillez, M. A., \& Breitschwerdt, D. 2004b, Ap\&SS, 292, 207
\bibitem[Bakes \& Tielens(1994)]{bak94} Bakes, E. L. O., \& Tielens, A. G. G. M. 1994, ApJ, 427, 822
\bibitem[Ballesteros-Paredes et al.(1999)]{bal99} Ballesteros-Paredes, J., Hartmann, L., \& V\'{a}zquez-Semadeni, E. 1999, ApJ, 527, 285
\bibitem[Benzi et al.(1993)]{ben93} Benzi, R., Ciliberto, S., Tripiccione, R., Baudet, C., Massaioli, F., \& Succi, S. 1993, Phys. Rev. E, 48, 29
\bibitem[Biferale et al.(2004)]{bif04} Biferale, L., Lanotte, A. S., \& Toschi, F. 2004, Phys. Rev. Lett. 92, 194503
\bibitem[Bland-Hawthorn \& Maloney(1999)]{bla99} Bland-Hawthorn, J., Maloney, P. R. 1999, ApJ, 510, L33
\bibitem[Bland-Hawthorn \& Maloney(2002)]{bla02} Bland-Hawthorn, J., Maloney, P. R. 2002, ASP Conf., 240, 267
\bibitem[Boldyrev(2002)]{bol02} Boldyrev, S. 2002, ApJ, 569, 841
\bibitem[Boldyrev et al.(2002)]{bnp02} Boldyrev, S., Nordlund, \AA., Padoan, P. 2002, ApJ, 573, 678
\bibitem[Bonnazzola et al.(1987)]{bon87} Bonazzola, S., Falgarone, E., Heyvaerts, J., Perault, M., \& Puget, J. L. 1987, A\&A, 172, 293
\bibitem[Boulares \& Cox(1990)]{bou90} Boulares, A., \& Cox, D. P. 1990, ApJ, 365, 544
\bibitem[Brunt \& Mac Low(2004)]{bru04} Brunt, C. M., \& Mac Low, M.-M. 2004, 604, 196
\bibitem[Brinks \& Bajaja(1986)]{bri86} Brinks, E., \& Bajaja, E. 1986, A\&A, 169, 14
\bibitem[Camussi \& Benzi(1996)]{cas96} Camussi, R., \& Benzi, R. 1996, Phys. Fluids Letters, 9, 257
\bibitem[Chandrasekhar(1951)]{cha51} Chandrasekhar, S. 1951, Proc. Royal Soc. London A, 210, 26
\bibitem[Cioffi et al.(1988)]{cio88} Cioffi, D. F., McKee, C. F., \& Bertschinger, E. 1988, ApJ, 334, 252
\bibitem[Clarke \& Oey(2002)]{cla02} Clarke, C., \& Oey, M. S. 2002, MNRAS, 337, 1299
\bibitem[Clemens et al.(1988)]{cle88} Clemens, D. P., Sanders, D. B., \& Scoville, N. Z. 1988, ApJ, 327, 139
\bibitem[Cole et al.(2000)]{col00} Cole, S., Lacey, C. G., Baugh, C. M., \& Frenk, C. S. 2000, MNRAS, 319, 168
\bibitem[Colella \& Woodward(1984)]{col84} Colella, P. Woodward, P. R. 1984, J. Comp. Phys., 54, 174
\bibitem[Dahlburg et al.(1990)]{dal90} Dahlburg, J. P., Dahlburg, R. B., Gardner, J. H., \& Picone, J. M. 1990, Phys. Fluids A, 2, 1481
\bibitem[Dalgarno \& McCray(1972)]{dal72} Dalgarno, A., \& McCray, R. A. 1972, ARA\&A, 10, 375
\bibitem[Dettmar(1992)]{det92} Dettmar, R. J. 1992, Fund. Cosmic Phys., 15, 143
\bibitem[Dickey \& Lockman(1990)]{dic90} Dickey, J. M., \& Lockman, F. J. 1990, ARA\&A, 28, 215
\bibitem[Domg\"{o}rgen \& Mathis(1994)]{dom94} Domg\"{o}rgen, H., \& Mathis, J. S. 1994, 428, 647
\bibitem[Dove \& Shull(1994)]{dov94} Dove, J. B., \& Shull,  J. M. 1994, ApJ, 430, 222
\bibitem[Draine(1978)]{dra78} Draine, B. T. 1978, ApJS, 36, 595
\bibitem[Dziourkevitch et al.(2004)]{dzi04} Dziourkevitch, N., Elstner, D., \& R\"{u}diger, G. 2004, A\&A, 423, 29
\bibitem[Elmegreen(2002)]{elm02} Elmegreen, B. G. 2002, ApJ, 577, 206
\bibitem[Elmegreen \& Scalo(2002)]{elm04} Elmegreen, B. G., \& Scalo, J. 2004, ARA\&A, 42, 211
\bibitem[Falkovich(1994)]{fal94} Falkovich, G. 1994, Phys. Fluids, 6, 1411
\bibitem[Ferri\`{e}re et al.(1991)]{fer91} Ferrière, K. M., Mac Low, M.-M. \& Zweibel, E. G. 1991, ApJ, 375, 239
\bibitem[Ferri\`{e}re(1995)]{fer95} Ferri\`{e}re, K. M. 1995, ApJ, 441, 281
\bibitem[Ferri\`{e}re(1998)]{fer98} Ferri\`{e}re, K. M. 1998, ApJ, 503, 700
\bibitem[Field et al.(1969)]{fie69} Field, G. B., Goldsmith, D. W., \& Habing, H. J. 1969, ApJ, 155, L49
\bibitem[Frisch(1995)]{fri95} Frisch, U. 1995, Turbulence. The Legacy of A. N. Kolmogorov, Cambridge: Cambridge University Press
\bibitem[Fryxell et al.(2000)]{fry00} Fryxell, B., Olson, K., Ricker, P., Timmes, F. X., Zingale, M. et al. 2000, ApJS, 131, 273
\bibitem[Fujita et al.(2003)]{fuj03} Fujita, A., Martin, C. L., Mac Low, M.-M., \& Abel, T. 2003, ApJ, 599, 50
\bibitem[Gerritsen \& Icke(1997)]{get97} Gerritsen, J. P. E., \& Icke, V. 1997, A\&A, 325, 972
\bibitem[Habing(1968)]{hab68} Habing, H. J. 1968, Bull. Astron. Inst. Netherlands, 19, 421
\bibitem[Haverkorn et al.(2004)]{hav04} Haverkorn, M., Gaensler, B. M., McClure-Griffiths, N. M., Dickey, John M., \& Green, A. J. 2004, 609, 776
\bibitem[Heckman et al.(2000)]{hec00} Heckman, T. M., Lehnert, M. D., Strickland, D. K., \& Armus, L. 2000, ApJS, 129, 493
\bibitem[Heckman(2001)]{hec01} Heckman, T. M. 2001, ASP Conf. Proc., Vol. 240, ed. J. E. Hibbard et al. (ASP, San Francisco), 345
\bibitem[Heiles(1987)]{hei87} Heiles, C. 1987, ApJ, 315, 555
\bibitem[Heiles(1990)]{hei90} Heiles, C. 1990, ApJ, 354, 483
\bibitem[Heiles(2001)]{hei01} Heiles, C. 2001, ApJ, 551, L105
\bibitem[Heiles \& Troland(2003)]{hei03} Heiles, C., \& Troland, T. H. 2003, ApJ, 586, 1067
\bibitem[Heiles \& Troland(2005)]{hei05} Heiles, C., \& Troland, T. H. 2005, ApJ, 624, 773
\bibitem[Heitsch et al.(2001)]{hei01} Heitsch, F., Mac Low, M.-M., \& Klessen, R. S. 2001, ApJ, 547, 280 (HMK01)
\bibitem[Hernquist \& Mihos(1995)]{her95} Hernquist, L., \& Mihos, J. C. 1995, ApJ, 448, 41
\bibitem[Hollenbach \& McKee(1979)]{hol79} Hollenbach, D. \& McKee, C. F. 1979, ApJS, 41, 555
\bibitem[Joung \& Mac Low(2005)]{jou05} Joung, M. K. R., \& Mac Low, M.-M. 2005, in preparation (Paper II)
\bibitem[Katz(1992)]{kat92} Katz, N. 1992, ApJ, 391, 502
\bibitem[Kennicutt et al.(1988)]{ken88} Kennicutt, R. C., Jr., Edgar, B. K., \& Hodge, P. W. 1988, ApJ, 337, 761
\bibitem[Kim et al.(2001)]{kim01} Kim, J., Balsara, D., \& Mac Low, M.-M. 2001, J. Kor. Astron. Soc., 34, 333
\bibitem[Kleiner \& Dickman(1984)]{kle84} Kleiner, S. C., \& Dickman, R. L. 1984, ApJ, 286, 255
\bibitem[Klessen(1998)]{kle98} Klessen, R. S. 1998, Ph.D. thesis, Max-Planck-Institut f\"{u}r Astronomie, Heidelberg
\bibitem[Klessen et al.(2000)]{kle00} Klessen, R. S., Heitsch, F., \& Mac Low, M.-M. 2000, ApJ, 535, 887 (KHM00)
\bibitem[Korpi et al.(1999)]{kor99} Korpi, M. J., Brandenburg, A., Shukorov, A., Tuominen, I., \& Nordlund, \AA. 1999, ApJ, 514, L99
\bibitem[Koyama \& Inutsuka(2004)]{koy04} Koyama, H., \& Inutsuka, S. 2004, ApJ, 602, L25
\bibitem[Kraichnan(1967)]{kra67} Kraichnan, R. H. 1967, Phys. Fluids 10, 1417
\bibitem[Kritsuk et al.(2004)]{kri04} Kritsuk, A. G., Norman, M. L., \& Padoan, P. 2004 (astro-ph/0411626)
\bibitem[Krumholz \& McKee(2005)]{kru05} Krumholz, M., \& McKee, C. F. 2005, ApJ, 630, 250
\bibitem[Kuijken \& Gilmore(1989)]{kui89} Kuijken, K., \& Gilmore, G. 1989, MNRAS, 239, 605
\bibitem[Larson(1981)]{lar81} Larson, R. B. 1981, MNRAS, 194, 809
\bibitem[Lazarian \& Esquivel (2003)]{laz03} Lazarian, A., \& Esquivel, A. 2003, ApJ, 592, L37
\bibitem[Li et al.(2003)]{li03} Li, Y., Klessen, R. S., \& Mac Low, M.-M. 2003, ApJ, 592, L975
\bibitem[Li et al.(2005)]{li05} Li, Y., Mac Low, M.-M., Klessen, R. S. 2005, submitted to ApJ (astro-ph/0501022)
\bibitem[Lithwick \& Goldreich(2001)]{li01} Lithwick, Y. \& Goldreich, P. 2001, ApJ, 562, 279
\bibitem[Lockman(2002)]{loc02} Lockman, F. J. 2002, ApJ, 580, L47
\bibitem[Mac Low \& McCray(1988)]{mac88} Mac Low, M.-M., \& McCray, R. 1988, ApJ, 324, 776
\bibitem[Mac Low et al.(1989)]{mac89} Mac Low, M.-M., McCray, R., \& Norman, M. L. 1989, ApJ, 337, 141
\bibitem[Mac Low et al.(1998)]{mac98} Mac Low, M.-M., Klessen, R. S., Burkert, A., \& Smith, M. D. 1998, Phys. Rev. Lett. 80, 2754
\bibitem[Mac Low(1999)]{mac99} Mac Low, M.-M. 1999, ApJ, 524, 169
\bibitem[Mac Low \& Ossenkopf(2000)]{mac04} Mac Low, M.-M., \& Ossenkopf, V. 2000, A\&A, 353, 339
\bibitem[Mac Low \& Klessen(2004)]{mac04} Mac Low, M.-M., \& Klessen, R. S. 2004, Rev. Mod. Phys., 76, 125
\bibitem[Mac Low et al.(2005)]{mac05} Mac Low, M.-M., Balsara, D. S., Kim, J., \& Avillez, M. A. 2005, ApJ, 626, 864
\bibitem[McCray \& Snow(1979)]{mcc79} McCray, R., \& Snow, T. P., Jr. 1979, ARA\&A, 17, 213
\bibitem[McKee \& Ostriker(1977)]{mck77} McKee, C. F., \& Ostriker, J. P. 1977, ApJ, 218, 148
\bibitem[McKee(1990)]{mck90} McKee, C. F. 1990, The Evolution of the Interstellar Medium, ASP Conf. Ser. 12, ed. L. Blitz (ASP, San Francisco), 55
\bibitem[McKee \& Williams(1997)]{mck97} McKee, C. F., \& Williams, J. P. 1997, ApJ, 476, 144
\bibitem[Miller \& Scalo(1979)]{mil79} Miller, G. E., \& Scalo, J. M. 1979, ApJS, 41, 513
\bibitem[Navarro \& White(1993)]{nav93} Navarro, J. F., \& White, S. D. M. 1993, MNRAS, 265, 271
\bibitem[Norman \& Ferrara(1996)]{nor96} Norman, C. A., \& Ferrara, A. 1996, ApJ, 467, 280
\bibitem[Ossenkopf \& Mac Low(2002)]{oss02} Ossenkopf, V., \& Mac Low, M.-M. 2002, A\&A, 390, 307
\bibitem[Ostriker \& McKee(1988)]{ost88} Ostriker, J. P., \& McKee, C. F. 1988, Rev. Mod. Phys., 60, 1
\bibitem[Padoan \& Nordlund(1999)]{pad99} Padoan, P., \& Nordlund, \AA. 1999, ApJ, 526, 279
\bibitem[Padoan et al.(1997a)]{pad97a} Padoan, P., Jones, B. J. T., \& Nordlund, \AA. 1997, ApJ, 474, 730
\bibitem[Padoan et al.(1997b)]{pad97b} Padoan, P., Nordlund, \AA., \& Jones, B. J. T. 1997, MNRAS, 299, 145
\bibitem[Padoan et al.(2003)]{pad03} Padoan, P., Boldyrev, S., Langer, W., \& Nordlund, \AA. 2003, ApJ, 583, 308
\bibitem[Parravano et al.(2003)]{par03} Parravano, A., Hollenbach, D. J., \& McKee, C. F. 2003, ApJ, 584, 797
\bibitem[Passot et al.(1988)]{pas88} Passot, T., Pouquet, A., \& Woodward, P. R. 1988, A\&A, 197, 228
\bibitem[Passot \& V\'{a}zquez-Semadeni(1998)]{pas98} Passot, T., \& V\'{a}zquez-Semadeni, E. 1998, Phys. Rev. E, 58, 4501
\bibitem[Pettini et al.(2001)]{pet01} Pettini, M., Shapley, A. E., Steidel, C. C. et al. 2001, ApJ, 554, 981
\bibitem[Pettini et al.(2002)]{pet02} Pettini, M., Rix, S. A., Steidel, C. C. et al. 2002, ApJ, 569, 742
\bibitem[Piontek \& Ostriker(2005)]{pio05} Piontek, R. A., \& Ostriker, E. C. 2005, ApJ, 629, 849
\bibitem[Porter et al.(1992)]{por92} Porter, D. H., Pouquet, A., \& Woodward, P. R. 1992, Phys. Rev. Lett. 68, 3156
\bibitem[Porter et al.(1998)]{por98} Porter, D. H., Woodward, P. R., \& Pouquet, A. 1998, Phys. Fluid, 10, 237
\bibitem[Press et al.(1992)]{pre92} Press, W. H., Teukolsky, S. A., Vetterling, W. T., \& Flannery, B. P. 1992, Numerical Recipes in FORTRAN: The Art of Scientific Computing, Cambridge: Cambridge University Press 
\bibitem[Reynolds(1991)]{rey91} Reynolds, R. J. 1991, The Interstellar Disk-Halo Connection in Galaxies, IAU Symp. No. 144, ed. H. Bloemen (Kluwer, Dordrecht), 67
\bibitem[Rosen et al.(1993)]{ros93} Rosen, A., Bregman, J. N., \& Norman, M. L. 1993, ApJ, 413, 137
\bibitem[Rosen \& Bregman(1995)]{ros95} Rosen, A., \& Bregman, J. N. 1995, ApJ, 440, 634
\bibitem[Sasao(1973)]{sas73} Sasao, T. 1973, Pub. Astron. Soc. Japan, 25, 1
\bibitem[Scalo et al.(1998)]{sca98} Scalo, J., V\'{a}zquez-Semadeni, E., Chappell, D., \& Passot, T. 1998, 504, 835
\bibitem[Sellwood \& Balbus(1999)]{sel99} Sellwood, J. A. \& Balbus, S. A. 1999, ApJ, 511, 660
\bibitem[Shapiro \& Field(1976)]{sha76} Shapiro, P. R., \& Field, G. B. 1976, ApJ, 205, 762
\bibitem[Shapley et al.(2003)]{sha03} Shapley, A. E., Steidel, C. C., Pettini, M., \& Adelberger, K. L. 2003, ApJ, 588, 65
\bibitem[She \& Leveque(1994)]{she94} She, Z., \& Leveque, E. 1994, Phys. Rev. Lett. 72, 336
\bibitem[She \& Waymire(1995)]{sw94} She, Z., \& Waymire, E. C. 1995, Phys. Rev. Lett., 74, 262
\bibitem[Shull \& Saken(1995)]{shu95} Shull, J. M., \& Saken, J. M. 1995, ApJ, 444, 663
\bibitem[Slavin \& Cox(1993)]{sla93} Slavin, J. D., \& Cox, D. P. 1993, ApJ, 417, 187
\bibitem[Slyz et al.(2005)]{sly05} Slyz, A. D., Devriendt, J. E. G., Bryan, G., \& Silk, J. 2005, MNRAS, 356, 737
\bibitem[Somerville \& Primack(1999)]{som99} Somerville, R. S., \& Primack, J. R. 1999, MNRAS, 310, 1087
\bibitem[Spaans \& Norman(1997)]{spa97} Spaans, M., \& Norman, C. 1997, ApJ, 483, 87
\bibitem[Stone et al.(1998)]{sto98} Stone, J. M., Ostriker, E. C., \& Gammie, C. F. 1998, ApJ, 508, L99
\bibitem[Strickland \& Stevens(2000)]{str00} Strickland, D., \& Stevens, I. 2000, MNRAS, 314, 511
\bibitem[Stutzki et al.(1998)]{stu98} Stutzki, J., Bensch, F., Heithausen, A., Ossenkopf, V., \& Zielinsky, M. 1998, A\&A, 336, 697
\bibitem[Sutherland \& Dopita(1993)]{sut93} Sutherland, R. S., \& Dopita, M. A. 1993, 88, 253
\bibitem[Tammann et al.(1994)]{tam94} Tammann, G. A., L\"{o}ffler, W., \& Schr\"{o}der, A. 1994, ApJS, 92, 487
\bibitem[Thacker \& Couchman(2001)]{tha01} Thacker, R. J., \& Couchman, H. M. P. 2001, ApJ, 555, L17
\bibitem[Toomre(1964)]{too64} Toomre, A. 1964, ApJ, 139, 1217
\bibitem[V\'{a}zquez-Semadeni \& Gazol(1995)]{vaz95} V\'{a}zquez-Semadeni, E., \& Gazol, A. 1995, A\&A, 303, 204
\bibitem[V\'{a}zquez-Semadeni et al.(1996)]{vaz96} V\'{a}zquez-Semadeni, E., Passot, T., \& Pouquet, A. 1996, ApJ, 473, 881
\bibitem[V\'{a}zquez-Semadeni et al.(2000)]{vaz00} V\'{a}zquez-Semadeni, E., Gazol, A., \& Scalo, J. 2000, ApJ, 540, 271
\bibitem[V\'{a}zquez-Semadeni et al.(2005)]{vaz05} V\'{a}zquez-Semadeni, E., Kim, J., \& Ballesteros-Paredes, J. 2005, ApJ, 630, L49
\bibitem[Wada \& Norman(2001)]{wad01} Wada, K., \& Norman, C. A. 2001, ApJ, 547, 172
\bibitem[Wakker(2001)]{wak01} Wakker, B. P. 2001, ApJS, 136, 463
\bibitem[Wood et al.(2004)]{woo04} Wood, K., Mathis, J. S., \& Ercolano, B. 2004, MNRAS, 348, 1337
\bibitem[Wolfire et al.(1995)]{wol95} Wolfire, M. G., Hollenbach, D., McKee, C. F., Tielens, A. G. G. M., \& Bakes, E. L. O. 1995, ApJ, 443, 152
\bibitem[Wolfire et al.(2003)]{wol03} Wolfire, M. G., McKee, C. F., Hollenbach, D., \& Tielens, A. G. G. M. 2003, 587, 278
\bibitem[Zuckerman \& Palmer(1974)]{zuc74} Zuckerman, B., \& Palmer, P. 1974, ARA\&A, 12, 279
\end{thebibliography}
\end{document}